%% file: main.tex
\documentclass[lettersize,journal]{IEEEtran}

\usepackage{amsmath,amsfonts,amssymb}
\usepackage{array}
\usepackage{makecell}
\usepackage{booktabs}
\usepackage{cite}
\usepackage{float}
\usepackage{placeins}
\usepackage{graphicx}
\usepackage{stfloats}
\usepackage{longtable}
\usepackage{multirow}
\usepackage[caption=false,font=normalsize,labelfont=sf,textfont=sf]{subfig}
\usepackage{tabularx}
\usepackage{textcomp}
\usepackage{url}
\usepackage{newtxtext,newtxmath}

\usepackage[ruled,vlined,linesnumbered]{algorithm2e}
\SetKwInput{KwIn}{Input}
\SetKwInput{KwOut}{Output}

\begin{document}

\title{{\fontsize{24pt}{28pt}\selectfont
Revisiting and Expanding the IPv6 Periphery: Global-Scale Measurement and Security Analysis}\\[8pt]}

\author{Zixuan~Xie, Zitao~Yang, Shurui~Fang, Zhaoyang~Li, Wenxing~Xie, Nannan~Fu, Liangyu~Dong, and Xiang~Li%
\thanks{\textit{Zixuan Xie and Zitao Yang contributed equally to this work.}}%
\thanks{\textit{Corresponding authors: Liangyu Dong; Xiang Li.}}
\thanks{Zixuan Xie, Shurui Fang, Zhaoyang Li, Wenxing Xie, and Xiang Li are with Nankai University, Tianjin, China (e-mail: \{2312585, 2213459, 2312092, 2310500\}@mail.nankai.edu.cn; lixiang@nankai.edu.cn).}%
\thanks{Zitao Yang, Nannan Fu, and Liangyu Dong are with China Industrial Control Systems Cyber Emergency Response Team and also with the Key Laboratory of Industrial Information Security Perception and Evaluation Technology, Ministry of Industry and Information Technology, Beijing 100040, China (e-mail: \{iamyangzt, fxycic96, dongliangyu78\}@163.com).}}%

\markboth{IEEE Transactions on Information Forensics and Security,~under review}%
{Xie \MakeLowercase{\textit{et al.}}: Revisiting and Expanding the IPv6 Network Periphery}

\IEEEaftertitletext{\vspace{-2\baselineskip}}

\maketitle

\begin{abstract}
\input{main/00_abstract}

\end{abstract}


\input{main/01_introduction}

\input{main/02_revisit_overview}
\input{main/03_ipv6_network_periphery_discovery}

\input{main/04_ipv6_exposed_service}

\input{main/05_routing_loop_attack}
\input{main/06_discussion}
\input{main/07_related_work}
\input{main/08_conclusion}

\input{main/09_acknowledgement}

\bibliographystyle{IEEEtran}
\bibliography{ref}

\end{document}

%% file: main/00_abstract.tex
As IPv6 deployment accelerates, understanding the evolving security posture of network peripheries becomes increasingly important.
A DSN 2021 study first explored IPv6 network peripheries, but its scope was limited to three regions and is now outdated.
In this paper, we revisit and expand that work through a global-scale measurement and security analysis of IPv6 network peripheries.
To support efficient large-scale scanning, we propose a novel Response-Guided Prefix Selection (RGPS) strategy to identify high-value IPv6 prefixes for probing.
Our measurement covers 73 countries/regions and identifies over 281.9M active IPv6 network peripheries, including a 364.8\% increase over the 2021 baseline in India, China, and America.
%We further analyze exposed services and find that 2.5\% of reachable services remain at risk, with recurring software vulnerabilities reflected by CVE correlations.
We further analyze exposed services and find that 2.5\% of devices expose at least one measured service, with recurring software vulnerabilities indicated by version-to-CVE mappings.
As a case study of emerging exposed services, we design a Hierarchical LLM Exposure Verification (HLEV) framework to identify unauthorized-access risks in LLM deployment tools.
Finally, we revisit routing loop vulnerabilities and identify 4.5M loop-prone devices, showing that flawed routing behaviors remain widespread.
These findings indicate that although IPv6 adoption has surged, key security challenges persist across network peripheries.

%% file: main/01_introduction.tex
\section{Introduction}

\IEEEPARstart{O}{ver} the past few years, IPv6 deployment has progressed steadily, with growing support from both networks and end-host devices.
APNIC Labs reported that, as of October 2024, IPv6 adoption had reached about 40\% of the global Internet user base~\cite{APNICLabs}.
Meanwhile, the IPv6 routing system continues to expand: as of March 2026, 36,034 ASes were originating IPv6 prefixes, and the BGP table contained 246,752 advertised IPv6 prefixes~\cite{BGP}.
Google’s statistics further show that 46.82\% of users accessed Google services over IPv6 in February 2026, a 54.1\% increase compared with January 2021~\cite{googleipv6}.

IPv6 introduces a vast address space, revised address allocation strategies, and direct end-to-end device communication~\cite{deering2017internet}.
Despite its widespread adoption, its security implications--particularly at the network periphery--remain insufficiently understood.
Previous measurement efforts have been limited in scope, geography, or technical coverage, and thus cannot fully capture the global IPv6 security landscape.

A notable contribution in this domain is the 2021 DSN study~\cite{li2021fast}, which discovered IPv6 network periphery devices by leveraging common address allocation strategies and XMap.
It identified 52M active IPv6 periphery devices across twelve major ISPs and revealed two major risks: unintended service exposure and exploitable routing loops.
However, it focused only on China, India, and America, and its findings are now four years old.

Motivated by these limitations, this work conducts a renewed and broader measurement of IPv6 network peripheries.
We extend the geographic scope to dozens of regions across multiple continents and revisit service exposure and routing security in today’s IPv6 landscape.
Our goal is to assess how the IPv6 security posture has evolved, identify emerging risks, and provide updated evidence for mitigation.

\noindent
\textbf{Our Paper.}
We provide a global security assessment of IPv6 network peripheries with a focus on changes over the past four years.
We first introduce our measurement methodology, experimental design, and toolchain (Section~\ref{sec:overview}), and then perform large-scale scanning to quantify the growth and geographic distribution of IPv6 network periphery devices (Section~\ref{sec:PeripheryDeviceDiscovery}), analyze service exposure and an LLM deployment-tool case study (Section~\ref{sec:IPv6Service}), and revisit routing-loop vulnerabilities (Section~\ref{sec:RoutingLoop}).

To evaluate IPv6 network periphery exposure, we adopt a two-stage measurement strategy.
First, we generate target addresses within selected prefixes using randomized probing techniques optimized for sparse IPv6 space.
Second, we capture and analyze responses from active hosts to identify periphery devices, characterize exposed services, and detect routing anomalies.
We use high-performance tools such as XMap and ZGrab2 to ensure scalable and protocol-level measurements (Section~\ref{sec:overview}).

To identify IPv6 network periphery devices at a global scale, we conduct a measurement campaign across diverse geographical and topological regions.
Guided by Response-Guided Prefix Selection (RGPS), our scanning covers high-value IPv6 prefixes selected from BGP-announced space across 73 regions.
These prefixes are operated by 164 major ISPs and span all five RIRs.
Through this methodology, we identify 281.9M IPv6 network periphery devices and reveal their highly uneven global distribution.
For comparison, we revisit the 15 original ISP blocks in India, China, and America studied in 2021~\cite{li2021fast}, where discovered devices increase from 52.6M to 244.8M (Section~\ref{sec:PeripheryDeviceDiscovery}).

To assess IPv6 service exposure, we revisit commonly used services and protocols at the network periphery.
Our broader coverage enables a more comprehensive view of remaining exposure risks, and we find that 2.5\% of reachable services remain under exposure-related risks.
We further map observed service versions to known CVEs as potential vulnerability indicators using official vulnerability references~\cite{cvesearch}.
As a case study of emerging exposed services, we also apply HLEV to identify unauthorized-access risks in LLM deployment tools, showing how AI-facing services can be verified through layered evidence (Section~\ref{sec:IPv6Service}).

To evaluate routing-loop vulnerabilities, we revisit this long-standing issue through global-scale ICMPv6-based validation.
We reassess the prevalence of loop-prone devices, analyze their global distribution, compare the results with the 2021 baseline, and examine the vendor landscape of affected devices.
Overall, we identify 4.5M loop-prone devices globally, while the proportion in the original three main regions declines substantially compared with 2021, indicating improvement but also persistent regional and operational disparities (Section~\ref{sec:RoutingLoop}).

\begin{figure*}[t]
    \centering
    \includegraphics[width=\textwidth]{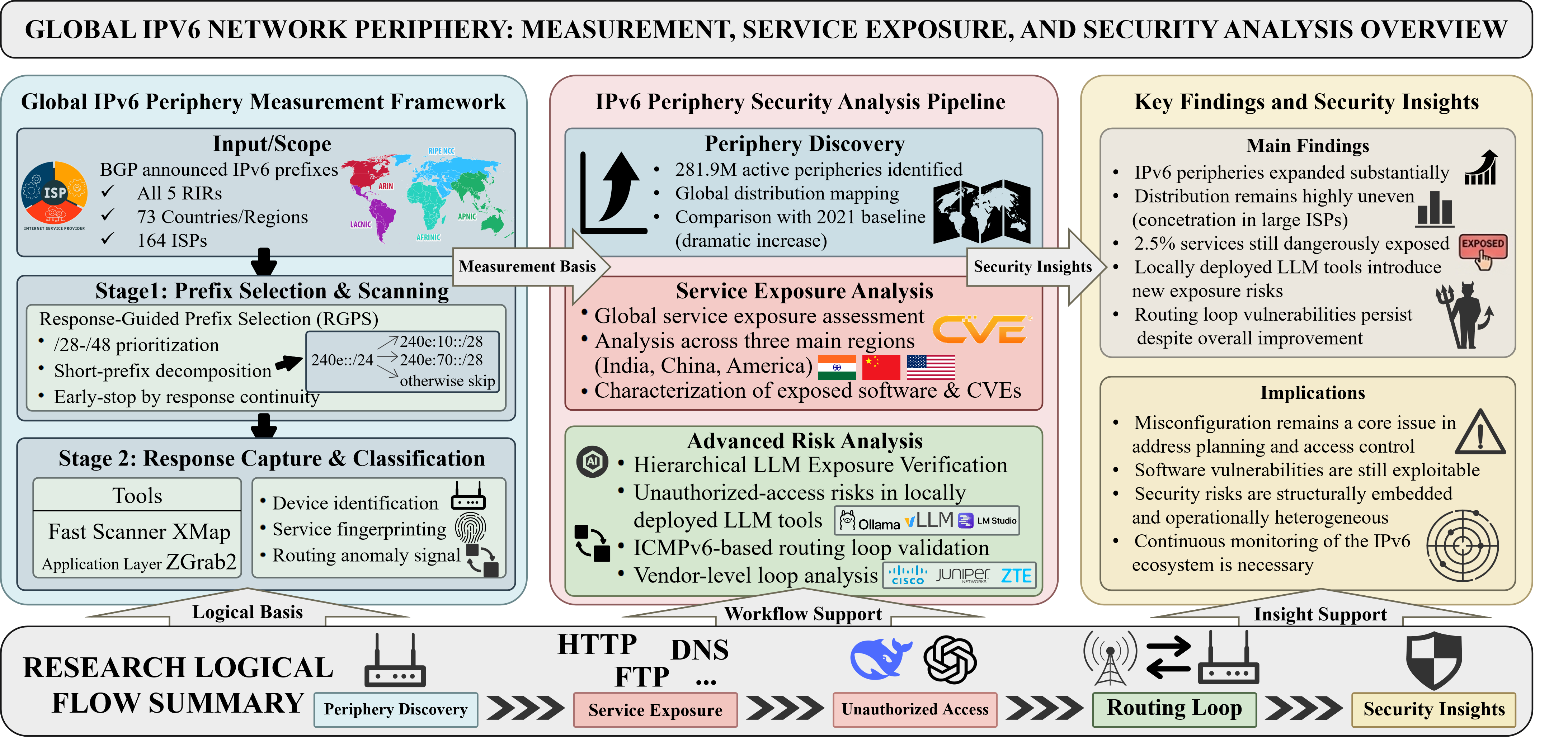}
    \caption{Global IPv6 Network Periphery: Measurement, Service Exposure, and Security Analysis Overview.}
    \label{fig:overview}
\end{figure*}

In summary, we revisit the current global landscape of IPv6 network peripheries, uncover persistent security risks, and derive mitigation insights.
Figure~\ref{fig:overview} summarizes our measurement framework, analysis pipeline, and main findings.

\noindent
\textbf{Our Contributions.}
We make the following contributions:

\begin{enumerate}
    \item We conduct a large-scale global measurement of IPv6 network peripheries across 73 regions and 164 major ISPs, identifying 281.9M active devices. To make Internet-scale probing over sparse IPv6 space practical, we introduce RGPS to prioritize high-value prefixes for efficient target discovery.

    \item We provide a comprehensive security analysis of the contemporary IPv6 periphery, covering exposed services, potential software vulnerabilities, an LLM deployment-tool case study using HLEV, and routing-loop vulnerabilities with ICMPv6 validation and vendor-level analysis.
    
    \item We build a scalable measurement and visualization platform for automated data collection, fingerprint analysis and longitudinal IPv6 security studies. If accepted, we will publicly release the framework and dataset to support reproducibility and continuous monitoring.
\end{enumerate}

%% file: main/02_revisit_overview.tex
\section{Revisit Overview}
\label{sec:overview}

This section presents the preparatory phase of our revisit, including the research goals, experiment design, and tools used for data acquisition and analysis.

\noindent
\textbf{Goals.}
This study aims to evaluate the global security posture of IPv6 network peripheries, with a focus on deployment trends, service exposure, and routing-loop vulnerabilities over the past four years.
We conduct large-scale measurements to characterize the growth and geographic distribution of IPv6 network periphery devices, identify unintended service exposure, and detect routing anomalies that reflect operational weaknesses.
Based on over 281.9M discovered devices, we provide an empirical foundation for understanding today’s IPv6 periphery security landscape and deriving data-driven mitigation insights.

\noindent
\textbf{Experiment Design.}
Our experiment follows a two-stage measurement pipeline:
\begin{enumerate}
    \item the prefix scanning and target generation stage;
    \item the response capture and classification stage.
\end{enumerate}
\noindent
In the first stage, we select high-value IPv6 prefixes from global BGP routing data~\cite{bgptoolkit} and generate probing targets within them.
To make large-scale scanning practical over the sparse IPv6 address space, this stage is guided by our Response-Guided Prefix Selection (RGPS) strategy, which focuses on prefixes with lengths ranging from /28 to /48 and further refines shorter prefixes when necessary.
These prefixes are operated by major ISPs across 73 regions with substantial IPv6 deployment.
In the second stage, we capture and analyze responses from active hosts to identify periphery devices, characterize exposed services, and detect potential vulnerabilities such as open administrative interfaces and routing anomalies.
For routing-loop detection, we monitor ICMPv6 messages~\cite{conta2006internet} and track Hop Limit variations to identify looping behaviors.
Finally, we conduct statistical analysis to assess the distribution and scale of identified devices across regions.

\noindent
\textbf{Tools.}
We primarily use XMap and ZGrab2 for our measurements~\cite{xmap,zgrab2}.
XMap is used for high-speed IPv6 probing and randomized traversal of selected prefix sets, enabling scalable discovery of active periphery devices.
ZGrab2 is then used for application-layer fingerprinting, service-version identification, and configuration retrieval on hosts discovered by XMap.
Together, they support broad IPv6 probing followed by detailed service characterization.

XMap is a high-performance IPv6 network probing framework derived from ZMap~\cite{durumeric2013zmap}.
It supports Internet-scale active measurements across the vast IPv6 address space through a generalized and fully permutable address-generation architecture.
Unlike tools restricted to specific prefixes or subnets~\cite{durumeric2013zmap,gont2020ipv6,masscan}, XMap enables randomized traversal of arbitrary IPv6 prefix sets through a redesigned address-generation module based on the GNU Multiple Precision (GMP) arithmetic library~\cite{gmp}.
This design converts IPv6 addresses into large-integer representations and applies permutation-based generation to produce mathematically valid address permutations within user-specified ranges.

XMap also adopts a modular architecture that integrates target-space processing, address randomization, protocol-specific probing, high-speed asynchronous packet I/O, and flexible result export.
Its protocol modules support IPv4/IPv6, TCP/UDP/ICMP, DNS, and application-layer behaviors, while its output pipeline allows selective fields such as addresses, ports, and payloads to be exported in TXT, CSV, or database formats~\cite{li2026xmap}.
These features make XMap suitable for large-scale IPv6 periphery discovery.

ZGrab2 is an application-layer scanner used after XMap identifies responsive hosts~\cite{zgrab2}.
In our workflow, it performs service fingerprinting and version identification for exposed services.
By combining XMap's scalable network-layer probing with ZGrab2's application-layer analysis, our measurement pipeline supports both broad IPv6 address-space exploration and fine-grained service characterization.

%% file: main/03_ipv6_network_periphery_discovery.tex
\section{IPv6 Network Periphery Discovery}
\label{sec:PeripheryDeviceDiscovery}

This section focuses on the discovery of IPv6 network periphery devices at a global scale.
Prior work analyzed only 15 ISP blocks from 12 major ISPs in India, China, and America~\cite{li2021fast}, which provided an important first look at IPv6 network peripheries but remained limited in scope.
To revisit and extend that measurement, we broaden the study to representative ISPs across all five RIRs and analyze the resulting global device distribution.
Specifically, we first describe how candidate prefixes are collected and selected for probing, then characterize the global distribution of the discovered devices, and finally compare our results with the 15 original blocks studied in 2021 to examine how deployment has evolved over the past four years.

\subsection{Prefix Selection}

For direct comparability with prior work, we retain the 15 ISP blocks studied in 2021~\cite{li2021fast}.
These blocks, drawn from 12 major ISPs in India, China, and America and categorized as broadband, mobile, and enterprise networks, serve as our longitudinal baseline.
Our preliminary scan of this subset reveals that even a single prefix from each block can already yield a device count comparable to earlier results.
Based on this observation, we select one prefix per block for the revisit to assess deployment changes in IPv6 network periphery devices.

To move beyond this limited baseline, we extend the measurement to all five RIRs and collect representative ISPs across 73 regions.
These regions are selected based on public country-level IPv6 deployment statistics from APNIC Labs~\cite{APNICIPv6Measurement} and Google~\cite{googleipv6}, with priority given to regions showing observable IPv6 capability and sufficient measurement samples.
This selection naturally emphasizes more active IPv6 deployments, but the same measurement procedure can be extended to other regions in the broader IPv6 Internet, where lower deployment density is expected to yield fewer discovered periphery devices.
We query the BGP Toolkit platform, a global BGP routing information source, using each ISP's name to identify its corresponding ASN~\cite{bgptoolkit}.
We then retrieve the announced IPv6 prefixes managed by these ISPs to construct the candidate pool for large-scale probing.

In the prefix scanning and target generation stage, we adopt a \emph{Response-Guided Prefix Selection} (RGPS) strategy to identify ``good'' prefixes for active scanning based on their likelihood of containing a high concentration of network periphery devices.
RGPS is orthogonal to recent IPv6 target-generation and adaptive scanning methods such as 6SENSE, AddrProbe, HMap, TNet, and 6Seeks~\cite{williams20246sense,cheng2025addrprobe,hou2026hmap,zhao2026tnet,yang20256seeks}.
While these methods focus on address-level target inference or generation, RGPS performs prefix-level filtering over BGP-announced prefixes to identify high-value scanning regions.
Algorithm~\ref{alg:prefix_selection} summarizes the workflow of the proposed RGPS strategy.
Under RGPS, the selection of a prefix as a ``good'' prefix is determined by the following criteria:

\begin{enumerate}
    \item \textbf{Prefix length between /28 and /48:}
    We prioritize prefixes with lengths ranging from /28 to /48.
    The upper bound of /48 follows common operational practice, where /48 is often regarded as the longest broadly routable IPv6 prefix~\cite{huston2006considerations}.
    The lower bound of /28 is chosen for efficiency, since scanning shorter prefixes at Internet scale is prohibitively expensive.

    \item \textbf{Decomposition of shorter prefixes:}
    Directly excluding prefixes shorter than /28 may miss many network periphery devices.
    Therefore, we first perform an exploratory scan on such prefixes, collect responsive source addresses from the \texttt{saddr} field, and derive active /28 sub-prefixes from the observed responses.
    Scanning these derived /28 sub-prefixes allows us to focus probing resources on responsive regions while approximating the coverage of the original shorter prefix.

    \item \textbf{Continuity of packet reception:}
    We monitor the continuity of responses during scanning to avoid wasting resources on low-density prefixes.
    If no responses are observed for $\tau$ consecutive minutes, the scan is terminated early.
    A smaller $\tau$ may prematurely stop sparse but responsive prefixes, whereas a larger $\tau$ wastes more resources on empty prefixes.
    In practice, we set $\tau=2$ minutes; since XMap probes the specified address space in a randomized order rather than sequentially~\cite{li2026xmap}, the absence of new responses for $\tau$ consecutive minutes usually indicates that few additional devices are likely to be discovered.
\end{enumerate}

\begin{algorithm}[t]
\caption{Response-Guided Prefix Selection}
\label{alg:prefix_selection}
\KwIn{Announced IPv6 prefix set $\mathcal{P}$; early-stop threshold $\tau$}
\KwOut{Selected good prefix set $\mathcal{G}$}

\BlankLine
initialize $\mathcal{G} \leftarrow \emptyset$\;

\ForEach{prefix $p \in \mathcal{P}$}{
    \uIf{$28 \leq |p| \leq 48$}{
        $\mathcal{C} \leftarrow \{p\}$\;
    }
    \uElseIf{$|p| < 28$}{
        perform an exploratory scan on $p$\;
        collect responsive source addresses from the \texttt{saddr} field\;
        derive active /28 sub-prefixes $\mathcal{C}$ from these responses\;
    }
    \Else{
        continue\;
    }

    \ForEach{candidate prefix $c \in \mathcal{C}$}{
        start scanning $c$\;
        \If{no responses are observed for $\tau$ consecutive minutes}{
            terminate the scan of $c$ early\;
        }
        \Else{
            $\mathcal{G} \leftarrow \mathcal{G} \cup \{c\}$\;
        }
    }
}
\Return{$\mathcal{G}$}\;
\end{algorithm}

Following these principles, we gather a large amount of IPv6 prefix information and conduct global-scale scanning of IPv6 network periphery devices.

\input{table/01_number_of_ipv6_network_peripheries}

\begin{figure}[t]
    \centering
    \includegraphics[width=1\linewidth]{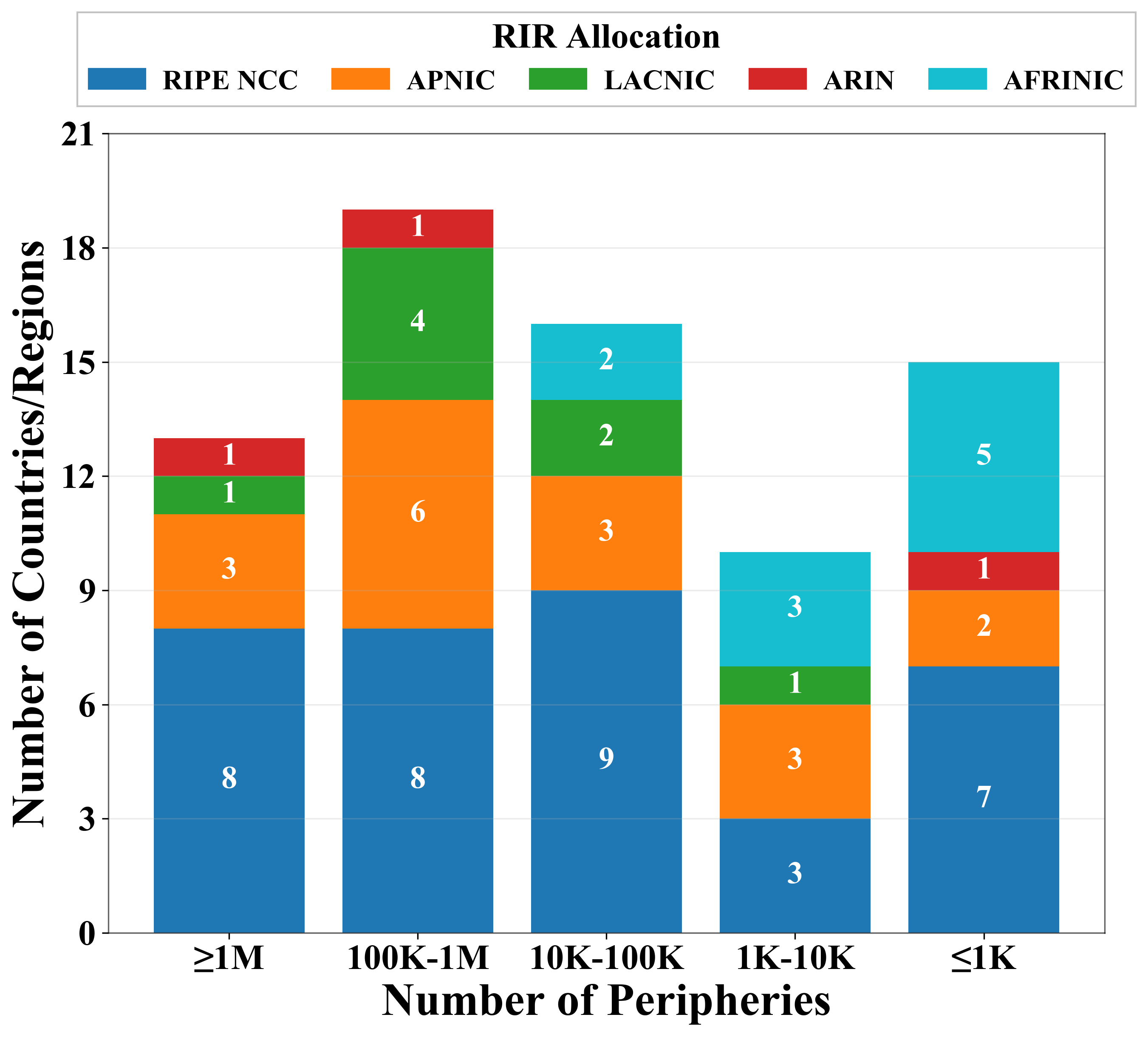}
    \caption{Distribution of Peripheries by Country/Region.}
    \label{fig:distribution_of_peripheries_by_country_region}
\end{figure}

\begin{figure}[t]
    \centering
    \includegraphics[width=1\linewidth]{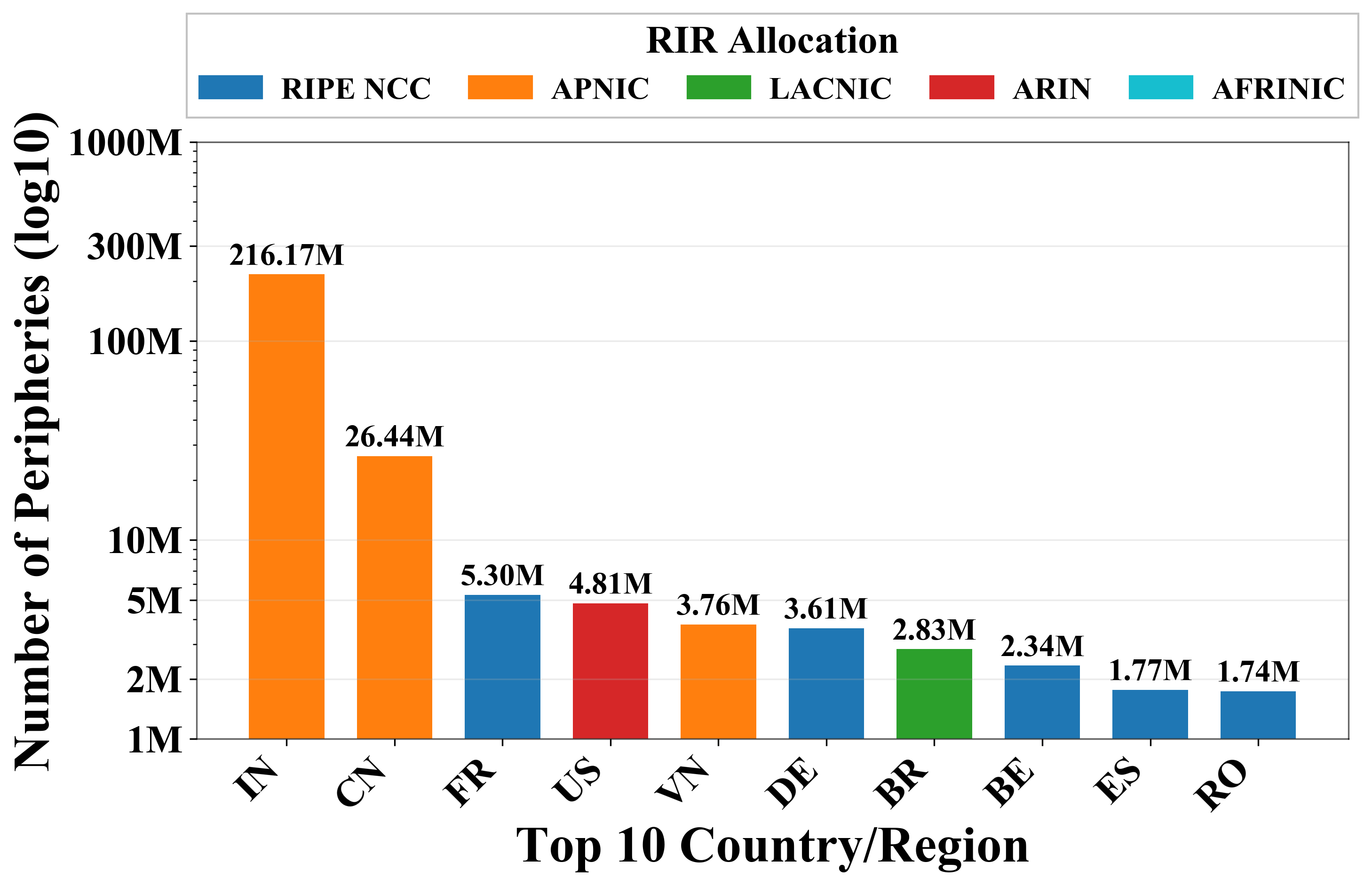}
    \caption{Top 10 Countries/Regions by Periphery Count.}
    \label{fig:top_10_countries_regions_by_periphery_count}
\end{figure}

\subsection{Global IPv6 Network Periphery Devices}

We identify a total of \textbf{281.9M} IPv6 network periphery devices through our scanning procedure.
Table~\ref{tab:number_of_ipv6_network_peripheries} shows that this footprint is highly uneven across RIRs.
APNIC alone contributes \textbf{249.51M} devices (\textbf{88.50\%}) across 17 regions and 43 ISPs, accounting for the overwhelming majority of all discovered devices.
By contrast, RIPE NCC covers the broadest regional span, with 35 regions and 63 ISPs, but contributes 22.05M devices (\textbf{7.82\%}), indicating a wider yet less concentrated deployment pattern.
ARIN and LACNIC contribute 5.76M (\textbf{2.04\%}) and 4.52M (\textbf{1.60\%}) devices, respectively, while AFRINIC currently contributes 87.40k (\textbf{0.03\%}) devices.

Figure~\ref{fig:distribution_of_peripheries_by_country_region} further shows a clear long-tailed distribution across the 73 measured regions.
Thirteen regions contain more than one million devices, while 19 fall into the 100K--1M range and 16 into the 10K--100K range; the remaining regions are distributed across the 1K--10K and $\leq$1K categories.
In terms of regional composition, RIPE NCC contributes the largest number of regions in several device-count bins, especially the medium-sized ones, whereas APNIC contributes fewer regions but dominates the highest-volume deployments.
This pattern suggests that global IPv6 periphery deployment is broad in geographic coverage, but uneven in device concentration.

The concentration becomes even clearer in Figure~\ref{fig:top_10_countries_regions_by_periphery_count}.
India (\textbf{216.17M}) and China (\textbf{26.44M}) are far ahead of the remaining regions, while the rest of the top 10 are all in the low-single-digit millions.
Overall, the global IPv6 periphery has expanded substantially, but its current distribution remains highly skewed toward a limited number of regions and networks.

\subsection{Devices in Three Main Regions}

For direct comparison with the 2021 study~\cite{li2021fast}, we revisit the 15 original ISP blocks in India, China, and America.
Based on our current scan, we observe a substantial increase in the number of IPv6 network periphery devices across these dominant blocks, reflecting rapid deployment and evolving allocation patterns in their corresponding access networks.

\input{table/02_ipv6_network_periphery_comparison}

Table~\ref{tab:ipv6_network_periphery_comparison} reports the detailed block-level statistics, including device counts and /64-prefix distributions.
The total number of detected devices across the 15 dominant IPv6 blocks has risen \textbf{from 52.6M to nearly 244.8M---an increase of 192.2M devices}.
India, in particular, demonstrates extraordinary growth: mobile devices under Bharti Airtel’s address space have surged from 22.5M to 205.6M, representing an approximately \textbf{9-fold increase}.
Similarly, broadband devices provided by BSNL have experienced more than a \textbf{75-fold growth}.

Among the 15 dominant blocks, nine exhibit substantial growth, with increases ranging from 49.6\% to 7440.4\%.
For the remaining six blocks, the observed decreases mainly reflect our reduced scanning scope, since the 2025 revisit probes only one representative prefix per block rather than the full block coverage used in 2021~\cite{li2021fast}.
Even so, the single-prefix results in these blocks still reach at least 8.6\% of the corresponding 2021 full-block measurements, suggesting that exhaustive scanning would likely reveal further growth.
Thus, our representative-prefix strategy captures the overall deployment trend while substantially reducing scanning cost.
Taken together, these results suggest that the overall trend across the revisited blocks is continued and often dramatic growth.

We also observe that the proportion of devices distributed across unique /64 IPv6 prefixes has generally decreased, excluding blocks with minimal variation (i.e., fluctuations below 1\%).
As the overall number of devices grows, a larger share is being assigned within the same /64 prefix.
This tendency may reflect addressing practices such as prefix reuse, denser subnet assignment, or shared access-network configurations.

Taken together, these longitudinal comparisons show that IPv6 deployment has continued to expand over the past four years, but the pace and structure of this growth differ substantially across RIRs, countries, ISPs, and access-network types.

%% file: table/01_number_of_ipv6_network_peripheries.tex
\begin{table}[t]
    \centering
    \small
    \caption{Number of IPv6 Network Peripheries}
    \label{tab:number_of_ipv6_network_peripheries}
    \begin{tabular*}{0.9\linewidth}{@{\extracolsep{\fill}}lcccc@{}}
    \toprule
    \textbf{RIR} & \textbf{Region} & \textbf{ISP} & \textbf{\#} & \textbf{\%} \\
    \midrule
    AFRINIC & 10 & 11 & 87.40k & 0.03\% \\
    APNIC & 17 & 43 & 249.51M & 88.50\% \\
    ARIN & 3 & 20 & 5.76M & 2.04\% \\
    LACNIC & 8 & 27 & 4.52M & 1.60\% \\
    RIPE NCC & 35 & 63 & 22.05M & 7.82\% \\
    \midrule
    \textbf{Total} & \textbf{73} & 164 & \textbf{281.92M} & \textbf{100\%} \\
    \bottomrule
    \end{tabular*}
\end{table}

%% file: table/02_ipv6_network_periphery_comparison.tex
\begin{table*}[!t]
  \centering
  \footnotesize
  \caption{IPv6 Network Periphery Devices and /64-Prefix Comparison (2021 vs. 2025)}
  \label{tab:ipv6_network_periphery_comparison}
  \begin{tabular*}{0.95\textwidth}{@{\extracolsep{\fill}} lllrrrrrrrrr @{}}
    \toprule
    \multirow{2}{*}{\textbf{Cty}} & \multirow{2}{*}{\textbf{Net}} & \multirow{2}{*}{\textbf{ISP}} & \multicolumn{4}{c}{\textbf{Last Hop}} & \multicolumn{5}{c}{\textbf{/64 prefix}} \\
    \cmidrule(lr){4-12}
          &       &       & \textbf{21\#} & \textbf{25\#} & \textbf{$\Delta$\#} & \textbf{\%incr} & \textbf{21\#} & \textbf{21\%} & \textbf{25\#} & \textbf{25\%} & \textbf{$\Delta$\%} \\
    \midrule
    \multirow{4}{*}{IN} & \multirow{2}{*}{Bb} & Reliance Jio & 3,365,175 & 2,301,020 & -1,064,155 & -31.6\% & 3,363,513 & 100.0 & 2,300,639 & 100.0 & 0.0 \\
          &       & BSNL & 2,404 & 181,271 & +178,867 & +7440.4\% & 2,276 & 94.7 & 181,159 & 99.9 & +5.2 \\
    \cmidrule{2-12}
          & \multirow{2}{*}{Mob} & Bharti Airtel & 22,542,690 & 205,676,491 & +183,133,801 & +812.3\% & 22,340,370 & 99.1 & 203,215,204 & 98.8 & -0.3 \\
          &       & Vodafone & 2,307,784 & 8,015,397 & +5,707,613 & +247.3\% & 2,307,672 & 100.0 & 8,014,921 & 100.0 & 0.0 \\
    \midrule
    \multirow{6}{*}{US} & \multirow{4}{*}{Bb} & Comcast & 87,308 & 1,641,709 & +1,554,401 & +1780.3\% & 5,694 & 6.5 & 8,773 & 0.5 & -6.0 \\
          &       & AT\&T & 740,141 & 110,221 & -629,920 & -85.1\% & 735,958 & 99.4 & 107,140 & 97.2 & -2.2 \\
          &       & Charter & 13,027 & 68,389 & +55,362 & +423.0\% & 1,573 & 12.1 & 2,665 & 3.9 & -8.2 \\
          &       & CenturyLink & 249,835 & 21,486 & -228,349 & -91.4\% & 233,298 & 93.4 & 16,094 & 74.9 & -18.5 \\
    \cmidrule{2-12}
          & Mob   & AT\&T & 1,734,506 & 331,637 & -1,402,869 & -80.9\% & 1,730,125 & 99.7 & 331,635 & 100.0 & +0.3 \\
    \cmidrule{2-12}
          & Ent   & Mediacom & 38,399 & 73,803 & +35,404 & +92.2\% & 516 & 1.3 & 490 & 0.7 & -0.6 \\
    \midrule
    \multirow{5}{*}{CN} & \multirow{3}{*}{Bb} & Telecom & 2,122,292 & 3,175,553 & +1,053,261 & +49.6\% & 2,100,034 & 99.0 & 3,148,221 & 99.1 & +0.1 \\
          &       & Unicom & 1,273,075 & 2,600,801 & +1,327,726 & +104.3\% & 1,272,540 & 100.0 & 2,597,763 & 99.9 & -0.1 \\
          &       & Mobile & 7,316,861 & 13,556,803 & +6,239,942 & +85.3\% & 7,315,713 & 100.0 & 13,529,403 & 99.8 & -0.2 \\
    \cmidrule{2-12}
          & \multirow{2}{*}{Mob} & Unicom & 3,696,275 & 2,488,610 & -1,207,665 & -32.7\% & 3,693,605 & 99.9 & 2,483,673 & 99.8 & -0.1 \\
          &       & Mobile & 7,193,972 & 4,616,008 & -2,577,964 & -35.8\% & 7,188,311 & 99.9 & 4,576,809 & 99.2 & -0.7 \\
    \midrule
    Total &       &       & 52,683,744 & 244,859,199 & +192,175,455 & +364.8\% & 52,291,198 & 99.3 & 240,514,589 & 98.2 & -1.1 \\
    \bottomrule
  \end{tabular*}
  
  \vspace{2mm}
\parbox{0.95\textwidth}{\scriptsize
  \textit{Notes:}
  Cty: country/region; Net: access-network type, where Bb, Mob, and Ent denote broadband, mobile, and enterprise networks, respectively.
  Last Hop reports the number of unique IPv6 network periphery devices, i.e., unique last-hop IPv6 addresses observed in each ISP block.
  /64 prefix reports the number and proportion of distinct /64 prefixes containing these last-hop addresses.
  21 and 25 denote the 2021 baseline and our 2025 revisit, respectively.
  \#: count; $\Delta$: 2025 vs. 2021 change; \%incr: relative increase of Last Hop count; $\Delta$\%: change in the /64-prefix proportion.
}
\end{table*}

%% file: main/04_ipv6_exposed_service.tex
\section{IPv6 Service Exposure}
\label{sec:IPv6Service}

% IPv6 significantly enhances global connectivity by providing an extensive address space ($2^{128}$) and improved protocol design, including simplified headers and native support for SLAAC and IPsec~\cite{deering2017internet,thomson2007ipv6,kent2005security}.
% However, these advantages also introduce new challenges for network management and security, particularly as IPv6 services become increasingly exposed on the Internet.

% These concerns are reflected in official deployment guidance.
% NIST notes that IPv6 deployment introduces security considerations that differ from IPv4~\cite{frankel2010guidelines}, while NSA warns that dual-stack operation can expand the attack surface and that transition tunnels may introduce unintended entry points~\cite{NSAIPv6}.

IPv6 significantly enhances global connectivity through a set of distinctive protocol features, including its extensive address space, simplified packet headers, native support for SLAAC, and mandatory support for IPsec~\cite{deering2017internet,thomson2007ipv6,kent2005security}.
At the same time, official deployment guidance has long recognized that these features also introduce new management and security challenges.
NIST notes that IPv6 deployment brings security considerations that differ from IPv4~\cite{frankel2010guidelines}, while NSA warns that dual-stack operation can expand the attack surface and that transition tunnels may create unintended entry points~\cite{NSAIPv6}.

Among these challenges, service exposure is particularly significant.
Empirical studies show that such risks already manifest at Internet scale.
Prior work on IPv6 network peripheries uncovered 52 million active periphery devices, including 4.7 million devices exposing unintended public services, 741 thousand open IPv6 DNS resolvers, and 1.3 million routers with Internet-accessible web management interfaces~\cite{li2021fast}.
More recently, 6SENSE conducted the first Internet-wide scanning-driven security analysis of IPv6 hosts and identified 81 thousand security-sensitive exposed devices together with at least 70 applicable CVEs~\cite{williams20246sense}.
Meanwhile, independent measurements show that large-scale IPv6 scanning is active in the wild and targets specific services and addresses~\cite{richter2022illuminating}.
These findings collectively suggest that IPv6 service exposure is no longer anecdotal, but a measurable and significant security problem.

Building on the IPv6 network periphery devices identified in Section~\ref{sec:PeripheryDeviceDiscovery}, we next investigate the global deployment of IPv6 services, with a focus on their exposure characteristics and vulnerability trends.
Specifically, this section first analyzes global IPv6 service exposure, then revisits the historical baseline in America, China, and India to examine how exposure has evolved in the same major ISP blocks, and finally assesses the software vulnerabilities commonly associated with the remaining exposed services.

\input{table/03_global_ipv6_network_periphery_service_exposure}

\subsection{Global IPv6 Service Exposure}

To characterize the security risks of the IPv6 network periphery devices discovered in Section~\ref{sec:PeripheryDeviceDiscovery}, we measure the exposure of commonly used IPv6 services at a global scale.
Unlike previous efforts four years ago, we significantly expand the scope of measurement to include major ISP-managed prefixes located in regions with mature IPv6 deployments.
For this work, we construct a \emph{Global IPv6 Periphery Fingerprint Database}, which serves as a foundational resource for understanding the evolving IPv6 landscape and its associated security implications (Table~\ref{tab:global_ipv6_network_periphery_service_exposure}).

\begin{figure}[t]
    \centering
    \includegraphics[width=1\linewidth]{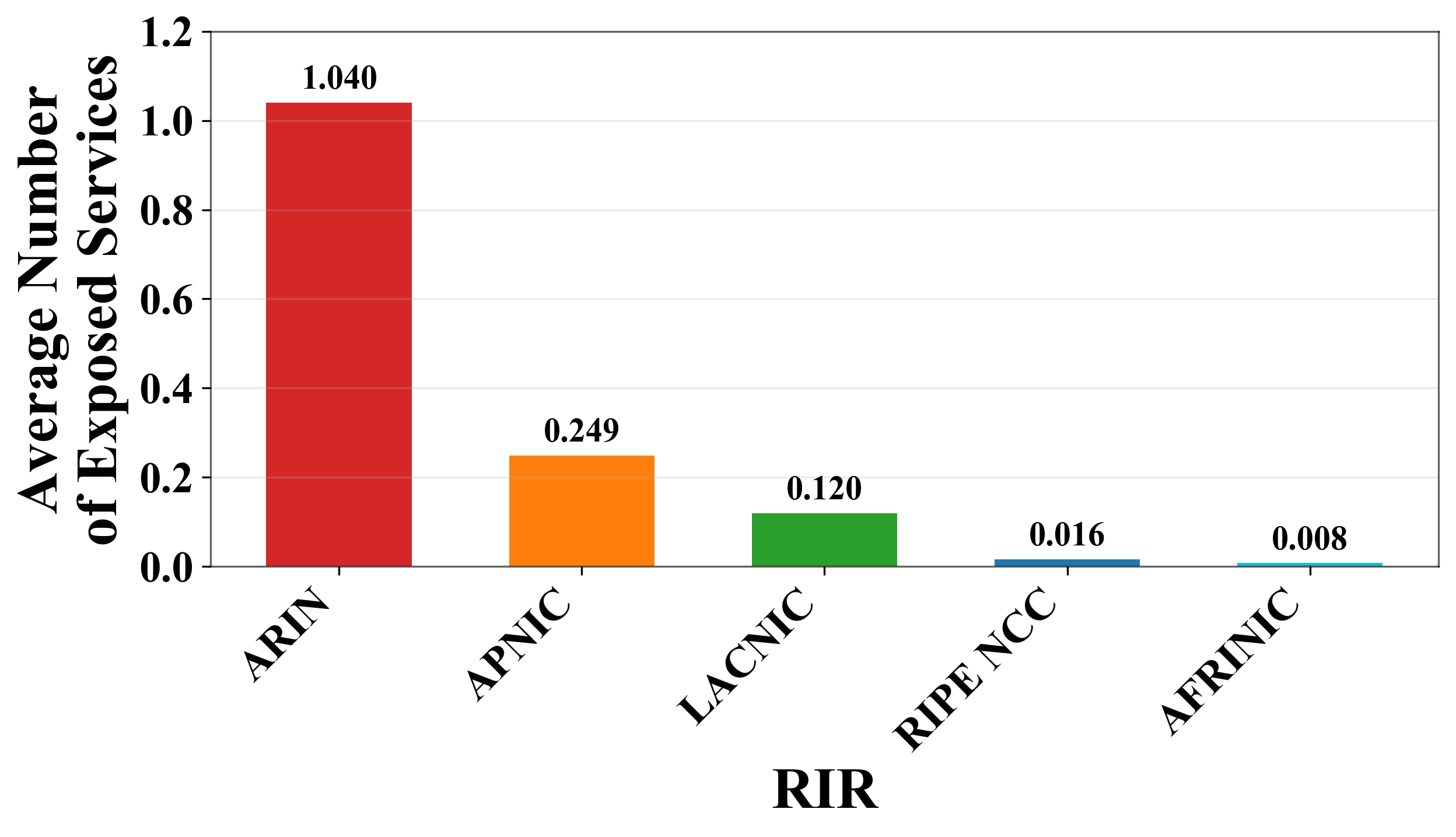}
    \caption{Average Number of Exposed Services per Periphery.}
    \label{fig:average_number_of_exposed_services_per_periphery}
\end{figure}

Among the discovered periphery devices, 2.5\% expose at least one of the measured services, with SSH (1.0\%) and HTTP/80 (0.7\%) remaining the most common risk points.
Despite a decrease in service exposure compared to 2021, approximately 7.1M devices remain exposed to insecure services.
At the RIR level, however, the exposure burden is highly uneven.
APNIC contributes 2.53M exposed devices, whereas ARIN contributes 3.70M, meaning that the largest absolute exposure is not observed in the largest deployment region but in the region with the highest exposure density.
This contrast is consistent with Figure~\ref{fig:average_number_of_exposed_services_per_periphery}: the average exposed service count per device reaches 1.040 in ARIN, far above APNIC (0.249), LACNIC (0.120), RIPE NCC (0.016), and AFRINIC (0.008).
In other words, APNIC dominates in deployment scale, while ARIN shows markedly higher exposure intensity on a per-device basis.

Table~\ref{tab:global_ipv6_network_periphery_service_exposure} further shows that the dominant exposed services also differ by RIR.
In APNIC, the main contribution comes from sheer scale, with HTTP/80 (1.08M) and FTP (952.2k) contributing substantially to the global total.
In ARIN, by contrast, exposure is concentrated in a smaller device population but with much higher densities, especially for SSH (47.6\%), HTTP/80 (8.3\%), and TLS (5.6\%).
LACNIC stands out for relatively high TELNET (5.4\%) and HTTP/80 (5.5\%) exposure ratios, suggesting that legacy management or embedded-device services remain visible there.
Overall, these results indicate that global IPv6 service exposure is shaped by two different forces: very large deployment scale in APNIC and markedly higher exposure density in ARIN and parts of LACNIC.

\subsection{Service Exposure in Three Main Regions}

This subsection focuses on unexpectedly exposed services in the same major ISP blocks revisited from India, America, and China, and analyzes how service exposure has changed in these main regions.

The current number of identified IPv6 network periphery devices in these blocks is nearly \textbf{five times higher} than it was four years ago.
Table~\ref{tab:ipv6_network_periphery_service_exposure_comparison} reports detailed block-level statistics on unexpectedly exposed services.

\input{table/04_ipv6_network_periphery_service_exposure_comparison}

In contrast to the sharp growth in device count, the total exposure across the eight key services has dropped substantially, indicating that large-scale IPv6 deployment has not been accompanied by a proportional increase in visible service exposure.

This reduction is evident across several major protocols.
Compared with 2021, HTTP/8080 decreases from 3.5M to 78k, HTTP/80 from 1.3M to 138k, DNS from 741k to 210k, and TLS from 144k to 31.2k.
In nine major IPv6 blocks (2, 3, 4, 5, 6, 7, 10, 11, and 12), the growth in device count has not resulted in a proportional increase in service exposure.
This strongly suggests that many ISPs have introduced tighter filtering policies, such as firewalls and access control lists, during network expansion.

At the same time, the decline is not uniform across all services and blocks.
Some ISP blocks still show noticeable growth in selected protocols, implying that hardening has been service-specific rather than universal.
For example, Comcast (block 5) shows clear increases in HTTP/80, TLS, and HTTP/8080 exposure, while China Telecom (block 11) exhibits growth in several management- or legacy-oriented services, including FTP, TELNET, and HTTP/8080.
Bharti Airtel (block 3), despite its explosive device growth, still shows non-negligible increases in SSH and TLS exposure.
These patterns likely reflect differences in ISP configuration practices, device ecosystems, and operational priorities.

Overall, while the exposure surface in the three main regions has become narrower than in 2021, a considerable number of exposed services still persist, ranging from thousands to hundreds of thousands of devices depending on the protocol.
Particularly for DNS, FTP, and SSH/TELNET-related services, the remaining exposure continues to provide a meaningful basis for large-scale exploitation and therefore warrants continued defensive attention.

\subsection{Software Vulnerabilities Associated with Exposed Services}

Table~\ref{tab:observed_service_versions_and_associated_cve_counts} summarizes the observed software versions and their associated CVE counts.
Although the overall exposure of several protocols has declined, the software stacks behind the remaining exposed services still exhibit non-negligible security relevance.
We note that such exposure may be intentional in some deployments, and banner-to-CVE matching may overestimate real exploitability due to incomplete version strings, vendor-specific modifications, or backported patches.
Therefore, the CVE counts in Table~\ref{tab:observed_service_versions_and_associated_cve_counts} should be interpreted as potential vulnerability indicators rather than confirmed exploitable vulnerabilities.

\input{table/05_observed_service_versions_and_associated_cve_counts}

% Table~\ref{tab:observed_service_versions_and_associated_cve_counts} summarizes the observed software versions and their associated CVE counts.
% Although the overall exposure of several protocols has declined, the software stacks behind the remaining exposed services still exhibit substantial vulnerability burden.

\noindent
\textbf{HTTP.}
Many exposed devices use common software versions such as micro\_httpd and Jetty.
These implementations remain security-relevant not only because HTTP exposure is still widespread, but also because they are often associated with weak authentication, poor input validation, and configuration errors~\cite{zilberman2025ipv6, jettyadvisories}.
For example, CVE-2014-4927 exposes a buffer overflow in ACME micro\_httpd, while CVE-2025-1948 highlights uncontrolled memory allocation in Eclipse Jetty HTTP/2 servers.

\noindent
\textbf{SSH.}
The SSH ecosystem remains fragmented across multiple versions of Dropbear and OpenSSH rather than being dominated by a single implementation.
This fragmentation suggests that exposed SSH services are difficult to secure uniformly because different implementations introduce different attack surfaces.
For example, CVE-2025-47203 exposes command injection in Dropbear SSH dbclient, while CVE-2025-26466 highlights memory-exhaustion DoS in OpenSSH.

\noindent
\textbf{DNS.}
The widespread use of dnsmasq version 2.7x is particularly noteworthy.
For example, CVE-2020-37127 exposes a buffer-overflow DoS in dnsmasq-utils dhcp\_release, while CVE-2018-13897 highlights hostname leakage through dnsmasq DNS records on Qualcomm Snapdragon devices.
Those issues indicate that DNS exposure is still closely tied to long-lived and insufficiently maintained software.

\noindent
\textbf{FTP.}
While the overall exposure of FTP services is lower, many legacy systems continue to run old FTP software on FreeBSD or ASUS versions.
This shows that even a relatively smaller exposed population can still correspond to high-risk software.

Overall, the remaining exposed services still rely heavily on outdated software, misconfigurations, and weak authentication mechanisms.
Therefore, exposure reduction alone is insufficient: there remains a need to strengthen security development lifecycle management, improve input validation, adhere to the principle of least privilege, and conduct regular vulnerability assessments to reduce the potential exploitability of exposed IPv6 services.

\subsection{Case Study: Unauthorized Access to LLM Tools}\label{subsec:llm_case_study}

Building on the exposed-service analysis above, we further conduct a focused case study on unauthorized-access risks in locally deployed Large Language Model (LLM) tools.
This part is not intended as a standalone Internet-wide LLM measurement.
Instead, it serves as an additional case study of how emerging application services may become exposed on IPv6 network periphery devices.
When local LLM runtime APIs or WebUI interfaces are made reachable without proper protection, attackers may enumerate available models, issue inference requests, or access sensitive interaction data.
Therefore, this case study extends our service-exposure analysis from traditional protocols to an emerging class of AI-facing services.

It should be noted that IPv6 network periphery devices are mainly routers, CPEs, and gateway-like infrastructure, which are generally less likely to host LLM runtime services than end hosts or cloud servers.
Thus, observing only a small number of confirmed LLM exposures in this setting is expected and should not be interpreted as evidence that LLM tools are rarely exposed on the broader Internet.
Rather, our goal is to provide a lightweight verification methodology that can be applied to IPv6 periphery devices and further extended to broader IPv6 end-host spaces or even IPv4 deployments.

\noindent
\textbf{Verification Methodology.}
To identify exposed LLM deployment tools, we design a \emph{Hierarchical LLM Exposure Verification} (HLEV) framework.
HLEV progressively refines candidates through three stages: transport-layer reachability, application-layer verification, and model-level confirmation.
Algorithm~\ref{alg:hlev} summarizes the detailed workflow.

\begin{algorithm}[t]
\caption{Hierarchical LLM Exposure Verification (HLEV)}
\label{alg:hlev}
\KwIn{Candidate IPv6 periphery address set $\mathcal{A}$; default LLM service port set $\mathcal{P}$; application-layer response signatures; model-specific confirmation endpoints and rules}
\KwOut{Verified exposed LLM service set $\mathcal{E}$}

initialize $\mathcal{E} \leftarrow \emptyset$\;

\BlankLine
\textbf{Transport-Layer Reachability}\;
\ForEach{address $a \in \mathcal{A}$}{
    perform TCP SYN probing on $a$ over ports in $\mathcal{P}$ using XMap\;
    \If{SYN-ACK response is observed}{
        record $a$ as a Response~0 candidate\;
    }
}

\BlankLine
\textbf{Application-Layer Verification}\;
\ForEach{Response~0 candidate $a$}{
    dispatch HTTP requests with default service settings using ZGrab2\;
    analyze the returned HTTP status codes, headers, and payload signatures\;
    \If{$a$ matches tool-specific response traits}{
        record $a$ as a Response~1 verified service\;
    }
}

\BlankLine
\textbf{Model-Level Confirmation}\;
\ForEach{Response~1 verified service $a$}{
    query model-related interfaces using tool-specific confirmation rules\;
    extract identifiers and metadata from responses such as model lists or configuration outputs\;
    cross-reference returned metadata with known model repositories or toolchain signatures\;
    \If{model exposure is confirmed}{
        $\mathcal{E} \leftarrow \mathcal{E} \cup \{a\}$\;
        record the corresponding Response~2 evidence\;
    }
}

\Return{$\mathcal{E}$}\;
\end{algorithm}

\smallskip
\noindent\emph{Transport-Layer Reachability.}
HLEV first probes the default ports of representative local LLM services.
A TCP SYN-ACK response is treated only as transport-layer evidence of reachability, rather than as proof of LLM exposure.

\smallskip
\noindent\emph{Application-Layer Verification.}
For reachable candidates, HLEV sends HTTP requests using ZGrab2 and inspects returned status codes, headers, and payloads.
The goal is to identify tool-specific response traits rather than relying only on open ports.
Table~\ref{tab:probing_requests_and_valid_responses} summarizes the application-layer signatures used in this step~\cite{ollamaep,lmstudio,vllmserver,janserver,gpt4allserver,xinferenceapi}.

\smallskip
\noindent\emph{Model-Level Confirmation.}
Finally, HLEV queries model-related interfaces according to tool-specific rules and extracts model identifiers or configuration metadata when available.
Only candidates that pass this stage are treated as confirmed exposed LLM services.

\input{table/06_probing_requests_and_valid_responses}

\noindent
\textbf{Case Study Results.}
Applying HLEV to the discovered IPv6 network periphery devices, we confirm one exposed Ollama instance.
This limited exposure is consistent with our measurement target, since IPv6 network periphery devices are primarily gateway or routing infrastructure and are not expected to commonly host local LLM services.
Therefore, this result should not be interpreted as evidence that LLM exposure is negligible across the broader IPv6 Internet; rather, it shows that such exposure can already appear even within periphery-device address spaces.
The case study also demonstrates that simple port scanning is insufficient for LLM exposure analysis, as HLEV requires both application-layer and model-level evidence before confirming an exposed deployment.

\noindent
\textbf{Security Implications.}
The Ollama exposure is security-sensitive because an exposed LLM runtime API may allow remote users to enumerate models, inspect service metadata, or issue inference requests through API endpoints.
Ollama has also been associated with unauthorized-access risks when its local service is bound to non-local interfaces without proper protection~\cite{CNVD-2025-04094}.
This suggests that LLM deployment interfaces should be treated as sensitive management services, especially when tools are deployed with weak access-control settings.

% \noindent
% \textbf{Case Study Results}
% Applying HLEV to the discovered IPv6 network periphery devices, we confirm one exposed Ollama instance.
% Although this number is small, it is consistent with the nature of our measurement target: IPv6 network periphery devices are primarily gateway or routing infrastructure and are not expected to commonly host local LLM services.

% This result should therefore be interpreted conservatively.
% It does not imply that LLM exposure is negligible across the broader IPv6 Internet; rather, it shows that such exposure can already appear even within periphery-device address spaces.
% More importantly, the case study demonstrates that simple port scanning is insufficient for LLM exposure analysis.
% A port-level match may indicate only a reachable HTTP service, while HLEV requires application-layer and model-level evidence before confirming an exposed LLM deployment.

% \noindent
% \textbf{Security Implications.}
% The confirmed Ollama exposure is security-sensitive because exposed LLM runtime APIs may allow remote users to enumerate models, inspect running sessions, or issue inference requests through API endpoints.
% Ollama has also been associated with unauthorized-access risks when its local service is bound to non-local interfaces without proper protection~\cite{CNVD-2025-04094}.
% Therefore, even isolated cases deserve attention, especially when LLM tools are deployed with default or weak access-control settings.

\noindent
\textbf{Domain-Specific Mitigation.}
For local LLM deployment tools, mitigation should go beyond generic service hardening.
Operators should keep LLM runtimes bound to loopback interfaces whenever possible, avoid exposing default ports such as 11434 to the public Internet, and place remote access behind authenticated reverse proxies or VPNs.
They should also restrict access to model-management and inference endpoints, such as model-listing, session-inspection, and chat-generation APIs, because these interfaces may reveal deployed models or consume expensive computation resources.
For WebUI-based deployments, login protection should be mandatory, and API backends should not remain reachable independently of the WebUI authentication layer.
In short, LLM deployment interfaces should be treated as sensitive management services rather than ordinary web applications.

Overall, this case study complements the IPv6 service-exposure analysis by showing how emerging AI-facing services can be verified in a layered manner.
While the confirmed IPv6 periphery exposure is limited, the same HLEV workflow can be extended to broader IPv6 end-host spaces and IPv4 deployments, where local LLM services are more likely to appear.

%% file: table/03_global_ipv6_network_periphery_service_exposure.tex
\begin{table*}[t]
  \centering
  \footnotesize
  \renewcommand{\arraystretch}{1}
  \caption{Global IPv6 Network Periphery Service Exposure}
  \label{tab:global_ipv6_network_periphery_service_exposure}
  \setlength{\tabcolsep}{2.25pt}
  \begin{tabular*}{0.95\textwidth}{@{\extracolsep{\fill}} c*{18}{r} @{}}
    \toprule
    \multirow{2}{*}{\textbf{RIR}} & 
    \multicolumn{2}{c}{\textbf{DNS}} &
    \multicolumn{2}{c}{\textbf{NTP}} &
    \multicolumn{2}{c}{\textbf{FTP}} &
    \multicolumn{2}{c}{\textbf{SSH}} &
    \multicolumn{2}{c}{\textbf{TELNET}} &
    \multicolumn{2}{c}{\textbf{HTTP80}} &
    \multicolumn{2}{c}{\textbf{TLS}} &
    \multicolumn{2}{c}{\textbf{HTTP8080}} &
    \multicolumn{2}{c}{\textbf{Total}} \\
    \cmidrule(lr){2-19}
     & \textbf{\#} & \textbf{\%} & \textbf{\#} & \textbf{\%} & \textbf{\#} & \textbf{\%} & \textbf{\#} & \textbf{\%} & \textbf{\#} & \textbf{\%} & \textbf{\#} & \textbf{\%} & \textbf{\#} & \textbf{\%} & \textbf{\#} & \textbf{\%} & \textbf{\#} & \textbf{\%} \\
    \midrule
    APNIC    & 232.9k & 0.1 & 2.8k   & 0.0 & 952.2k & 0.4 & 35.1k  & 0.0 & 33.3k  & 0.0 & 1.08M  & 0.4 & 83.5k  & 0.0 & 112.2k & 0.0 & 2.53M  & 35.5 \\
    ARIN     & 5.2k   & 0.1 & 7.7k   & 0.1 & 49.4k  & 0.9 & 2.74M  & 47.6 & 21.6k  & 0.4 & 476.3k & 8.3 & 324.5k & 5.6 & 76.8k  & 1.3 & 3.70M  & 52.0 \\
    RIPE NCC & 63.5k  & 0.3 & 1.8k   & 0.0 & 38.0k  & 0.2 & 25.0k  & 0.1 & 45.3k  & 0.2 & 105.3k & 0.5 & 60.8k  & 0.3 & 4.5k   & 0.0 & 343.1k & 4.8 \\
    LACNIC   & 3.2k   & 0.1 & 2.4k   & 0.1 & 117    & 0.0 & 39.6k  & 0.9 & 245.5k & 5.4 & 249.8k & 5.5 & 3.2k   & 0.1 & 1.2k   & 0.0 & 541.9k & 7.6 \\
    AFRINIC  & 18     & 0.0 & 200    & 0.2 & 84     & 0.1 & 159    & 0.2 & 128    & 0.2 & 29     & 0.0 & 55     & 0.1 & 9      & 0.0 & 682    & 0.0 \\
    \midrule
    Total    & 304.8k & 0.1 & 14.9k  & 0.0 & 1.04M  & 0.4 & 2.84M  & 1.0 & 345.8k & 0.1 & 1.91M  & 0.7 & 472.1k & 0.2 & 194.7k & 0.1 & 7.12M  & 2.5 \\
    \bottomrule
  \end{tabular*}

  \vspace{2mm}
  \parbox{0.95\textwidth}{\scriptsize
    In the service columns (DNS, NTP, etc.), the percentage represents the proportion of devices within each RIR exposing that specific service.
    In the Total column, the percentage represents the proportion of exposed devices of each RIR relative to the global total exposed devices.
  }
\end{table*}

%% file: table/04_ipv6_network_periphery_service_exposure_comparison.tex
\begin{table*}[!t]
  \centering
  \caption{IPv6 Network Periphery Service Exposure Comparison (2021 vs. 2025)}
  \label{tab:ipv6_network_periphery_service_exposure_comparison}
  \resizebox{\textwidth}{!}{
    \begin{tabular}{c*{24}{r}}
    \toprule
    \multirow{2}[4]{*}{\textbf{ISP}} & \multicolumn{3}{c}{\textbf{DNS}} & \multicolumn{3}{c}{\textbf{NTP}} & \multicolumn{3}{c}{\textbf{FTP}} & \multicolumn{3}{c}{\textbf{SSH}} & \multicolumn{3}{c}{\textbf{TELNET}} & \multicolumn{3}{c}{\textbf{HTTP80}} & \multicolumn{3}{c}{\textbf{TLS}} & \multicolumn{3}{c}{\textbf{HTTP8080}} \\
    \cmidrule{2-25}          & \textbf{21\#} & \textbf{25\#} & \textbf{$\Delta$\#} & \textbf{21\#} & \textbf{25\#} & \textbf{$\Delta$\#} & \textbf{21\#} & \textbf{25\#} & \textbf{$\Delta$\#} & \textbf{21\#} & \textbf{25\#} & \textbf{$\Delta$\#} & \textbf{21\#} & \textbf{25\#} & \textbf{$\Delta$\#} & \textbf{21\#} & \textbf{25\#} & \textbf{$\Delta$\#} & \textbf{21\#} & \textbf{25\#} & \textbf{$\Delta$\#} & \textbf{21\#} & \textbf{25\#} & \textbf{$\Delta$\#} \\
    \midrule
    1     & 30.3k & 59    & -30.2k & 6     & 4     & -2    & 1     & 1     & 0     & 9     & 4     & -5    & 1     & 8     & +7    & 102   & 14    & -88   & 0     & 0     & 0     & 1.4k  & 0     & -1.4k \\
   
    \textbf{2} & 4     & 9     & +5    & 88    & 21    & -67   & 21    & 9     & -12   & 89    & 25    & -64   & 55    & 10    & -45   & 24    & 1     & -23   & 20    & 2     & -18   & 4     & 0     & -4 \\
    
    \textbf{3} & 36.6k & 5.7k  & -30.9k & 131   & 0     & -131  & 27    & 23    & -4    & 50    & 748   & +698  & 19    & 24    & +5    & 1k    & 19    & -981  & 0     & 3.4k  & +3.4k & 6.7k  & 1     & -6.7k \\
    
    \textbf{4} & 201   & 0     & -201  & 39    & 3     & -36   & 0     & 6     & +6    & 13    & 15    & +2    & 2     & 0     & -2    & 141   & 206   & +65   & 0     & 0     & 0     & 623   & 0     & -623 \\
    
    \textbf{5} & 9     & 98    & +89   & 290   & 46    & -244  & 5     & 4     & -1    & 13    & 128   & +115  & 50    & 6     & -44   & 54    & 5.7k  & +5.6k & 64    & 5.9k  & +5.8k & 319   & 2.1k  & +1.8k \\
    
    6     & 3.6k  & 7     & -3.6k & 320   & 2.7k  & +2.4k & 880   & 61    & -819  & 223   & 533   & +310  & 13    & 40    & +27   & 340   & 9     & -331  & 3.4k  & 87    & -3.3k & 0     & 0     & 0 \\

    \textbf{7} & 437   & 130   & -307  & 58    & 130   & +72   & 1     & 1     & 0     & 46    & 30    & -16   & 3     & 5     & +2    & 31    & 20    & -11   & 372   & 242   & -130  & 357   & 78    & -279 \\

    8     & 3.6k  & 6     & -3.6k & 14.9k & 2.8k  & -12.1k & 1k    & 93    & -907  & 1.9k  & 569   & -1.3k & 1.5k  & 76    & -1.4k & 38    & 8     & -30   & 3k    & 89    & -2.9k & 2     & 0     & -2 \\
    
    9     & 0     & 0     & 0     & 0     & 0     & 0     & 0     & 0     & 0     & 3     & 0     & -3    & 2     & 0     & -2    & 625   & 0     & -625  & 625   & 0     & -625  & 489   & 0     & -489 \\
    
    \textbf{10} & 93    & 54    & -39   & 129   & 155   & +26   & 14    & 2     & -12   & 1.2k  & 208   & -992  & 1.1k  & 162   & -938  & 2.6k  & 77    & -2.5k & 1.3k  & 489   & -811  & 55    & 18    & -37 \\
  
    \textbf{11} & 63.6k & 16.3k & -47.3k & 146   & 396   & +250  & 211   & 3.5k  & +3.3k & 335   & 501   & +166  & 240   & 1k    & +760  & 791   & 1.5k  & +709  & 51    & 20.7k & +20.7k & 7     & 947   & +940 \\
    
    \textbf{12} & 202k  & 186k  & -16k  & 76    & 951   & +875  & 35.8k & 956   & -34.8k & 20.5k & 471   & -20k  & 36.5k & 1.1k  & -35.4k & 211k  & 130k  & -81k  & 169   & 259   & +90   & 229k  & 74.7k & -155k \\
   
    \textbf{13} & 403k  & 652   & -402k & 19    & 14    & -5    & 139k  & 8     & -139k & 114k  & 7     & -114k & 140k  & 47    & -140k & 1M    & 86    & -1M   & 138k  & 24    & -138k & 3.3M  & 172   & -3.3M \\
    
    14    & 468   & 2     & -466  & 21    & 11    & -10   & 0     & 0     & 0     & 8     & 7     & -1    & 5     & 1     & -4    & 147   & 11    & -136  & 4     & 16    & +12   & 176   & 53    & -123 \\
  
    15    & 296   & 2     & -294  & 122   & 0     & -122  & 0     & 0     & 0     & 133   & 5     & -128  & 130   & 3     & -127  & 96    & 0     & -96   & 1     & 0     & -1    & 236   & 3     & -233 \\
    \midrule
    Total & 741k  & 210k  & -531k & 16.1k & 7.3k  & -8.8k & 177k  & 4.7k  & -172k & 139k  & 3.3k  & -135k & 180k  & 2.5k  & -177k & 1.3M  & 138k  & -1.16M & 144k  & 31.2k & -113k & 3.5M  & 78k   & -3.42M \\
    \bottomrule
    \end{tabular}%
    }
    
    \vspace{2mm}
\parbox{\textwidth}{\scriptsize India: 1=Reliance Jio\textsuperscript{b}, 2=BSNL\textsuperscript{b}, 3=Bharti Airtel\textsuperscript{m}, 4=Vodafone\textsuperscript{m}\quad US: 5=Comcast\textsuperscript{b}, 6=AT\&T\textsuperscript{b}, 7=Charter\textsuperscript{b}, 8=CenturyLink\textsuperscript{b}, 9=AT\&T\textsuperscript{m}, 10=Mediacom\textsuperscript{e}\quad China: 11=Telecom\textsuperscript{b}, 12=Unicom\textsuperscript{b}, 13=Mobile\textsuperscript{b}, 14=Unicom\textsuperscript{m}, 15=Mobile\textsuperscript{m}\quad \textsuperscript{b}:Broadband, \textsuperscript{m}:Mobile, \textsuperscript{e}:Enterprise.}
  \label{tab:exposurecomparison}%
\end{table*}%

%% file: table/05_observed_service_versions_and_associated_cve_counts.tex
\begin{table}[t]
  \centering
  \small
  \caption{Observed Service Versions and Associated CVE Counts}
  \label{tab:observed_service_versions_and_associated_cve_counts}
  \setlength{\tabcolsep}{3pt}
  \renewcommand{\arraystretch}{1.05}

  \begin{tabularx}{\columnwidth}{@{}c>{\raggedright\arraybackslash}Xr@{}}
    \toprule
    \textbf{Service} & \textbf{Top Software \& Version (\# device)} & \textbf{\# CVE} \\
    \midrule
    HTTP & micro\_httpd (126,922), Jetty (74,671), Xfinity (7,612), Boa (670), MiniWeb (461), AkamaiGHost (206) & 241 \\

    SSH & dropbear\_unknown (1,085), 2012.55 (161), 0.5x (148), OpenSSH\_7.x (281), 3.5p1 (87), 8.x (60) & 202 \\

    DNS & dnsmasq 2.7x (7,026), 2.3x (692), 2.8x (145), 2.4x (92), 2.9x (42) & 44 \\

    FTP & FreeBSD 6.00LS (85), ASUS RT/GT (97), vsFTPd (76), FRITZ!Box (12) & 154 \\
    \bottomrule
  \end{tabularx}
\end{table}

%% file: table/06_probing_requests_and_valid_responses.tex
\begin{table}[!htbp]
  \centering
  \small
  \caption{Probing Requests and Valid Responses}
  \label{tab:probing_requests_and_valid_responses}
  \setlength{\tabcolsep}{3pt}
  \renewcommand{\arraystretch}{1.5}
  \begin{tabular}{@{}p{1.55cm}p{0.85cm}p{3.45cm}p{2.45cm}@{}}
    \toprule
    \textbf{Tool} & \textbf{Port} & \textbf{Match 1} & \textbf{Match 2} \\
    \midrule
    Ollama 
      & 11434 
      & \makecell[l]{"body"="Ollama\\ running"} 
      & -- \\

    LM Studio 
      & 1234  
      & \makecell[l]{"body"="Unexpected\\ endpoint (GET /)"} 
      & -- \\

    GPT4All 
      & 4891  
      & \makecell[l]{"content-type"=\\ "application/x-empty"} 
      & \makecell[l]{grep "GPT4All"} \\

    Jan.ai 
      & 1337  
      & \makecell[l]{"location"="./static\\ /index.html"} 
      & -- \\

    vLLM  
      & 8000  
      & \makecell[l]{"body"="\{detail: Not\\ Found\}"} 
      & \makecell[l]{grep "vLLM"} \\

    Xinference 
      & 9997  
      & \makecell[l]{"location"="/ui/"} 
      & -- \\

    LobeChat 
      & 3210  
      & \makecell[l]{grep "lobechat"} 
      & -- \\
    \bottomrule
  \end{tabular}
\end{table}

%% file: main/05_routing_loop_attack.tex
\section{Routing Loop Vulnerabilities}
\label{sec:RoutingLoop}

In addition to the exposed-service analysis and the LLM exposure study in Section~\ref{sec:IPv6Service}, we further investigate routing loop vulnerabilities on the discovered IPv6 network periphery devices.
This section presents an active detection methodology based on ICMPv6 error messages, then analyzes the global distribution of routing loop vulnerabilities, revisits the 15 ISP blocks studied in 2021 for longitudinal comparison, and finally examines the vendor distribution of affected devices through service fingerprinting.
Overall, we identify 4,517,099 devices involved in routing loops, accounting for 1.6\% of all scanned devices.
At the global level, routing loop vulnerability rates vary substantially across regions; for the 15 historical ISP blocks, the affected proportion decreases from 11.04\% in 2021 to 1.43\% in our revisit, indicating that although the problem persists, its prevalence in the original baseline has declined markedly.

\subsection{Routing Loop Detection Methodology}

Our approach adopts an active probing mechanism based on ICMPv6 error messages to identify network periphery devices susceptible to IPv6 routing loops.
The core idea is to trigger forwarding behavior toward unassigned targets and then infer abnormal looping paths from the returned ICMPv6 diagnostics.

\noindent
\textbf{Response Capture and Classification.}
The detection logic centers on collecting and analyzing ICMPv6 \texttt{Time Exceeded} messages (Type~3), which indicate that a packet has traversed beyond the configured Hop Limit.
During probing, we manipulate the Hop Limit field in IPv6 headers to elicit feedback from intermediate routers.
All returned ICMPv6 messages are captured and parsed in accordance with RFC~4443~\cite{conta2006internet}, and each response is classified by type and code to distinguish ordinary unreachable behavior from loop-induced feedback:

\begin{enumerate}
    \item \textbf{Type=1 (Destination Unreachable):} Indicates that the destination network or address is unreachable, requiring further inspection of the Code field.
    \item \textbf{Type=3 (Time Exceeded):} Typically generated when the Hop Limit expires in transit, and therefore serves as the main signal for potential forwarding loops.
\end{enumerate}

In the initial probing phase, an ICMPv6 Echo Request is transmitted to a target located in an unassigned subnet with a default Hop Limit of 32.
This value is selected based on empirical evidence that 99\% of Internet paths involve fewer than 32 hops~\cite{beverly2018ip}, thereby reducing false negatives while limiting the amplification of potential loop effects.
Considering the continued expansion of the IPv6 routing ecosystem, we further validate this threshold using recent Route Views data~\cite{routeviews} and observe that the overwhelming majority of contemporary IPv6 routing paths still remain within 32 hops.
When a target device returns a Type=3 message, we retransmit the same packet with an incremented Hop Limit value (i.e., $h+2$).
If another Type=3 message is received, we confirm the presence of a routing loop, indicating that the probe is circulating between forwarding devices rather than progressing toward a valid destination.
Repeated probing is applied to reduce false positives and improve robustness.

\subsection{Global Routing Loop Vulnerability}

\input{table/07_routing_loop_distribution_of_peripheries}

\begin{figure}[t]
    \centering
    \includegraphics[width=1\linewidth]{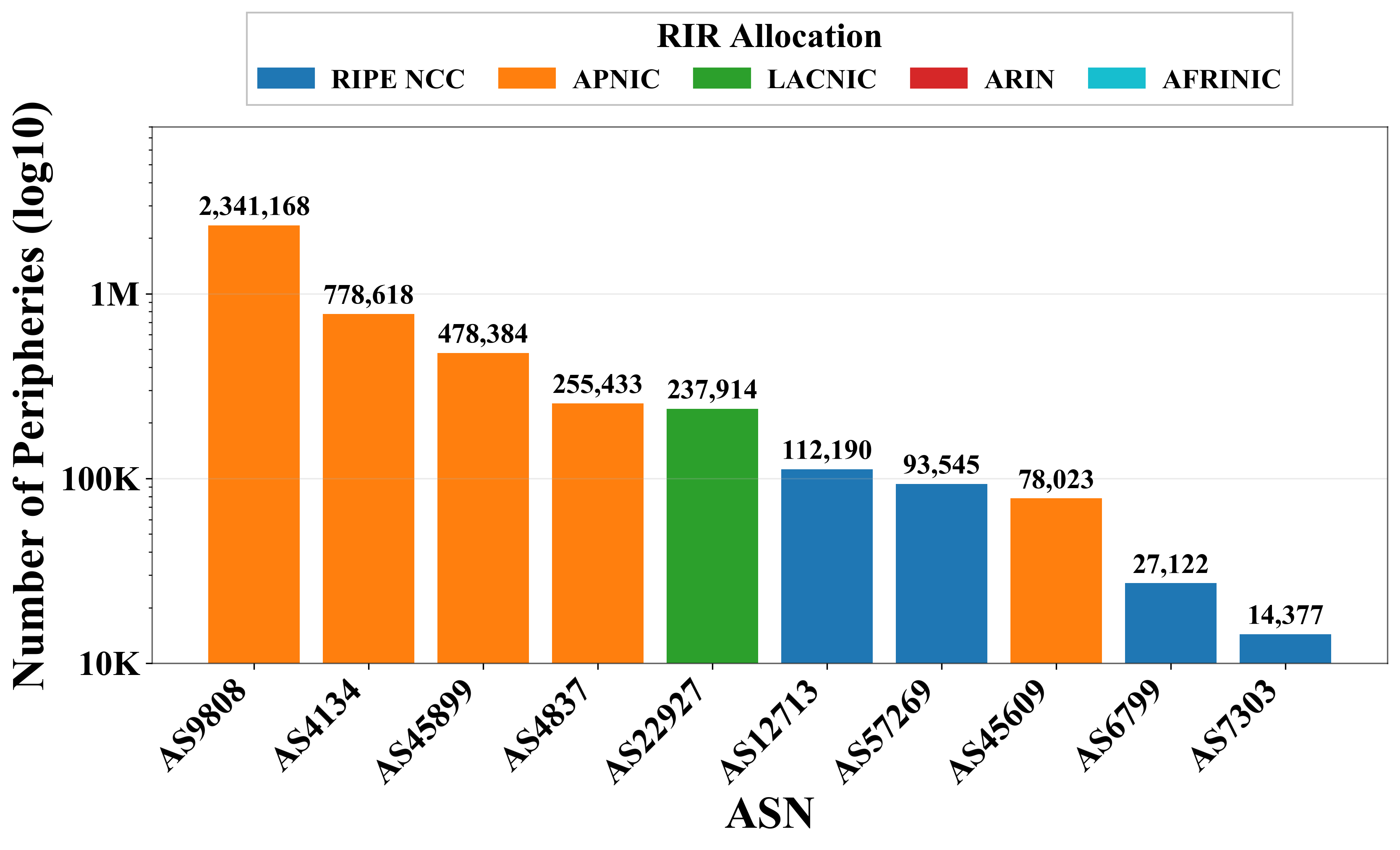}
    \caption{Top 10 ASNs by Routing Loop.}
    \label{fig:top_10_asns_by_routing_loop}
\end{figure}

\begin{figure}[t]
    \centering
    \includegraphics[width=1\linewidth]{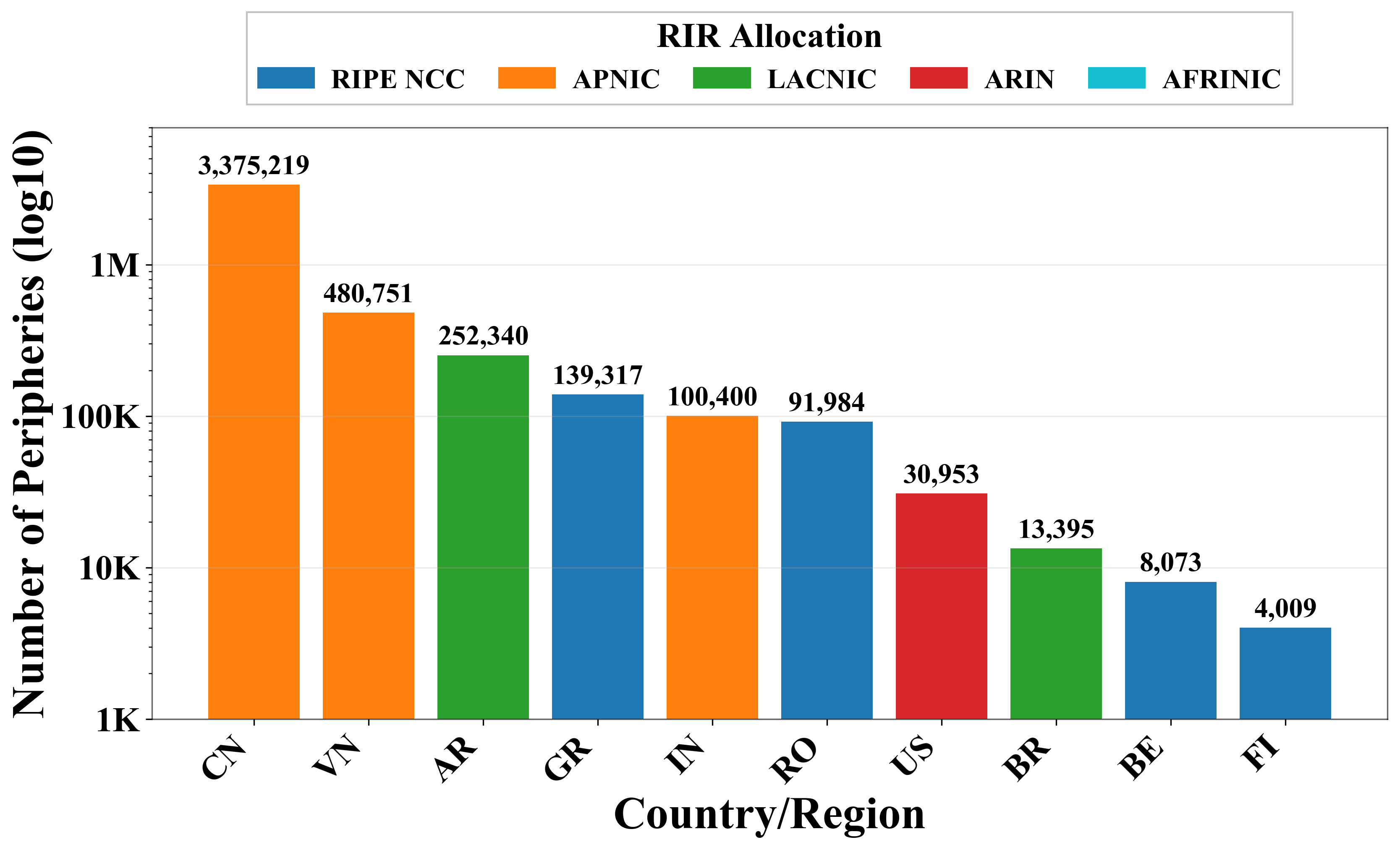}
    \caption{Top 10 Countries/Regions by Routing Loop.}
    \label{fig:top_10_countries_regions_by_routing_loop}
\end{figure}

On a global scale, the average routing loop vulnerability rate stands at 1.6\%.
Table~\ref{tab:routing_loop_distribution_of_peripheries} shows that this risk is highly uneven across regions.
APNIC contributes the largest number of affected devices, with 3.96M vulnerable devices out of 249.51M total devices, corresponding to a rate of 1.59\%.
By contrast, LACNIC has only 4.52M total devices but 0.27M affected ones, yielding the highest regional vulnerability rate at 6.01\%.
ARIN exhibits the lowest rate, at 0.54\%, together with the smallest number of affected devices among the large-deployment regions, suggesting comparatively stronger operational mitigation.
RIPE remains in the middle, with 0.25M affected devices and a rate of 1.15\%, while AFRINIC shows a low rate of 0.62\% but over a much smaller IPv6 deployment base.

This unevenness is further reflected in Figure~\ref{fig:top_10_asns_by_routing_loop} and Figure~\ref{fig:top_10_countries_regions_by_routing_loop}.
The vulnerable population is strongly concentrated in a small number of ASNs and countries rather than being evenly distributed across all deployments.
In Figure~\ref{fig:top_10_asns_by_routing_loop}, the largest four ASNs all belong to APNIC, with AS9808 alone contributing 2,341,168 affected devices and AS4134 contributing another 778,618.
Figure~\ref{fig:top_10_countries_regions_by_routing_loop} shows a similarly skewed pattern at the country level: China contributes 3,375,219 affected devices, far ahead of the second-ranked region, Vietnam, with 480,751.
Together, these results indicate that routing loop vulnerabilities remain a structurally concentrated problem, driven by a limited number of large networks and regions rather than by uniform global weakness.

\subsection{Routing Loop Vulnerability in Three Main Regions}

For direct comparison with the 2021 study, we revisit the same 15 IPv6 address blocks in India, China, and America.

\input{table/08_peripheries_under_routing_loop_attack}

Table~\ref{tab:peripheries_under_routing_loop_attack} summarizes the block-level comparison results.
The overall number of devices under routing loop attack decreases from 5.8M to 3.5M, while the corresponding proportion drops sharply from 11.04\% to 1.43\%.
This large reduction suggests that mitigation efforts introduced after the earlier disclosure have had a substantial effect on the original baseline.

The reduction is particularly evident in the major Chinese broadband blocks.
China Telecom (block 11), China Unicom (block 12), and China Mobile (block 13) all show large absolute declines, with the corresponding vulnerability ratios decreasing by 15.22, 69.02, and 35.72 percentage points, respectively.
These three blocks account for most of the improvement in the revisit and indicate that large-scale remediation can significantly reduce routing loop exposure even in very large access networks.

At the same time, the improvement is not uniform across all blocks.
Several US broadband blocks continue to exhibit elevated or even increased vulnerability ratios, most notably block 6 (AT\&T) and block 8 (CenturyLink), whose proportions rise to 10.44\% and 57.66\%, respectively.
These exceptions show that routing loop vulnerabilities remain dynamic and operationally heterogeneous.
Since our 2025 revisit uses representative prefixes within each historical block, individual block-level changes should be interpreted with some caution; nevertheless, the aggregate decline across the baseline is large enough to indicate meaningful overall improvement rather than a purely measurement-driven effect.

\subsection{Periphery Device Vendor Information}

To characterize routing loop vulnerabilities at the vendor level, we perform service fingerprinting on devices already confirmed to be involved in routing loops and extract device metadata to infer associated hardware manufacturers.

\noindent
\textbf{Service Fingerprinting and Vendor Identification.}
We analyze the packet contents of eight exposed services and find that HTTP on ports 80 and 8080, FTP on port 21, and Telnet on port 23 are particularly likely to disclose vendor-specific information.
For devices confirmed to be involved in routing loops, we use ZGrab2 to rescan common ports (HTTP/80, HTTP/8080, FTP/21, and Telnet/23).
We then extract metadata such as \texttt{Server}, \texttt{WWW-Authenticate}, \texttt{X-Powered-By}, page title (from HTTP), and service banners (from FTP/Telnet) to infer device vendors.

\begin{figure}[t]
    \centering
    \includegraphics[width=1\linewidth]{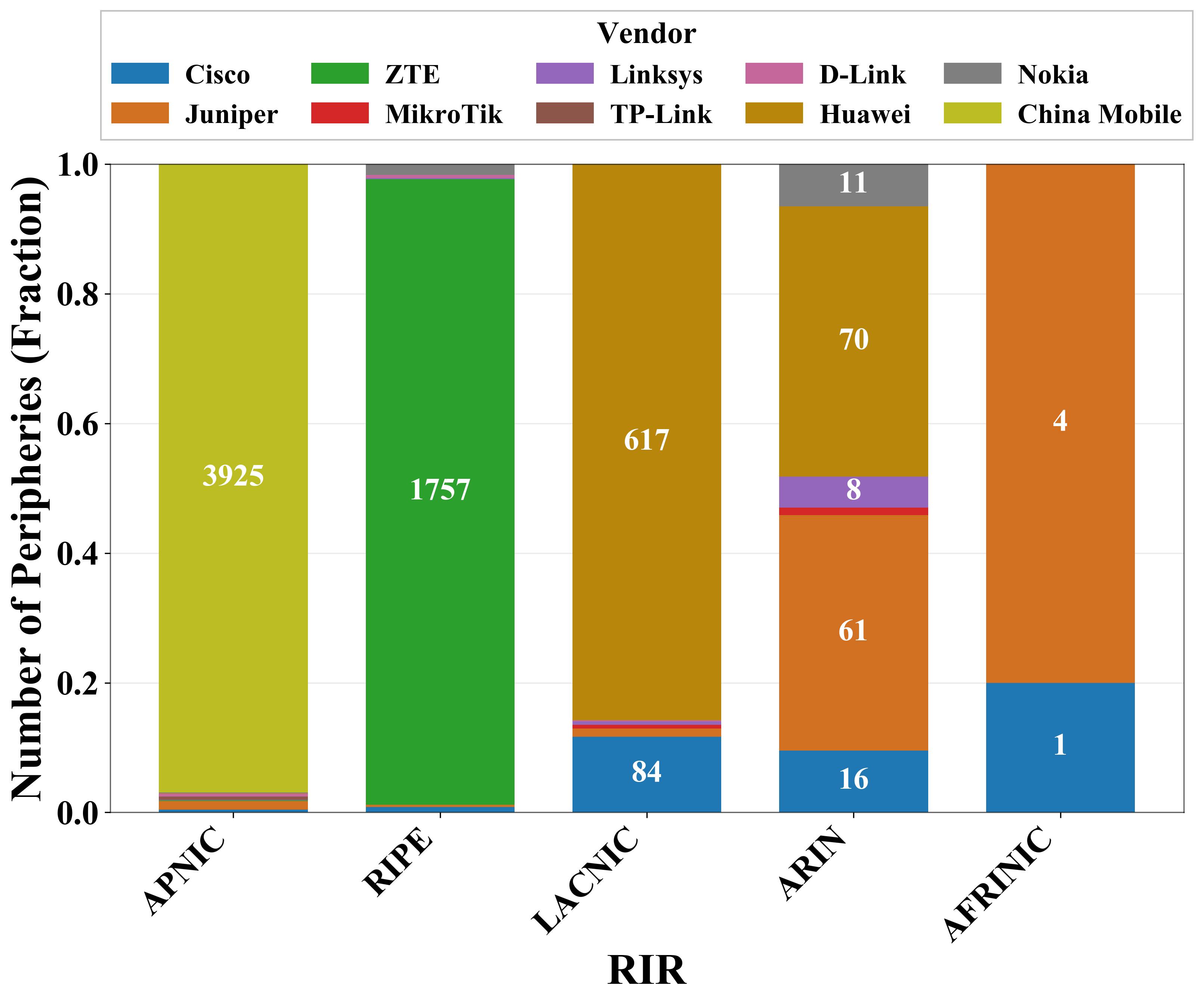}
    \caption{Top 10 Vendors by Routing Loop Periphery Count.}
    \label{fig:top_10_vendors_by_routing_loop_periphery_count}
\end{figure}

\textit{We find that 22 vendors' CPE devices exhibit routing loop vulnerabilities.}
Figure~\ref{fig:top_10_vendors_by_routing_loop_periphery_count} presents the top ten most common vendors associated with routing loop vulnerabilities and their corresponding RIRs, with each region normalized to fractions to highlight proportional structure.
The resulting pattern is again highly uneven.
APNIC is dominated by China Mobile, contributing 3,925 identified devices, whereas RIPE is overwhelmingly dominated by ZTE with 1,757 devices.
LACNIC is primarily associated with Huawei, which contributes 617 devices, while ARIN shows a comparatively more diverse mix, including Huawei (70), Juniper (61), Cisco (16), Linksys (8), and Nokia (11).
AFRINIC contains only a very small number of identified vendors, mainly Juniper and Cisco.

These results suggest that routing loop vulnerabilities are not tied to a single vendor family, but instead arise across multiple hardware ecosystems and operational environments.
At the same time, the strong dominance of a few vendors within particular RIRs indicates that regional market structure and deployment preferences materially influence the observed attack surface.
Cisco and Juniper appear across multiple regions, highlighting that the problem is globally distributed rather than limited to one local vendor ecosystem.

To address routing loop attacks, network equipment vendors have introduced a range of mitigation mechanisms, including hop-limit enforcement~\cite{junipertransportip}, loop suppression techniques such as Split Horizon and Poison Reverse~\cite{ciscorip,ciscoxr12krip}, and protocol- and control-plane safeguards in modern routing systems, such as OSPFv3~\cite{coltun2008ospf} and BGP AS-path loop detection~\cite{juniperbgploops}, as well as, more recently, AI-assisted dynamic route optimization.

%% file: table/07_routing_loop_distribution_of_peripheries.tex
\begin{table}[t]
  \centering
  \small
  \caption{Routing Loop Distribution of Peripheries}
  \label{tab:routing_loop_distribution_of_peripheries}
  \begin{tabular}{cccc}
    \toprule
    \textbf{RIR} & \textbf{\#Vulnerable} & \textbf{\#Total} & \textbf{\%} \\
    \midrule
    AFRINIC & 0.54k & 87.40k & 0.62\% \\
    APNIC   & 3.96M & 249.51M  & 1.59\% \\
    ARIN    & 31.06k & 5.76M  & 0.54\% \\
    LACNIC  & 0.27M & 4.52M  & 6.01\% \\
    RIPE    & 0.25M & 22.05M & 1.15\% \\
    \midrule
    \textbf{Total} & \textbf{4.51M} & \textbf{281.92M} & \textbf{1.60\%} \\
    \bottomrule
  \end{tabular}
\end{table}

%% file: table/08_peripheries_under_routing_loop_attack.tex
\begin{table}[!htbp]
  \centering
  \caption{Peripheries Involved in Routing Loops}
  \label{tab:peripheries_under_routing_loop_attack}
  \setlength{\tabcolsep}{3pt}
  \begin{tabular}{cccrrrrrr}
    \toprule
    \multirow{2}[2]{*}{\textbf{Cty}} & \multirow{2}[2]{*}{\textbf{Net}} & \multirow{1}[4]{*}{\textbf{ISP}} & \multicolumn{6}{c}{\textbf{Last Hop}} \\
    \cmidrule{4-9}
          &       &       & \textbf{21 \#} & \textbf{21\%} & \textbf{25 \#} & \textbf{25\%} & \textbf{$\Delta$\#} & \textbf{$\Delta$\%} \\
    \midrule
    \multirow{4}[4]{*}{IN} & \multirow{2}[2]{*}{Bb} & 1  & 8,606   & 0.26  & 830      & 0.04  & -7,776    & -0.22 \\
    \cmidrule{3-9}
          &       & 2  & 324     & 13.48 & 21,481   & 11.85 & +21,157   & -1.63 \\
    \cmidrule{2-9}
          & \multirow{2}[2]{*}{Mob} & 3  & 29,135  & 0.13  & 78,023   & 0.04  & +48,888   & -0.09 \\
    \cmidrule{3-9}
          &       & 4  & 207     & 0.01  & 66       & 0.00  & -141      & -0.01 \\
    \midrule
    \multirow{6}[6]{*}{US} & \multirow{4}[4]{*}{Bb} & 5  & 31      & 0.04  & 422      & 0.03  & +391      & -0.01 \\
    \cmidrule{3-9}
          &       & 6  & 1,598   & 0.22  & 11,510   & 10.44 & +9,912    & +10.22 \\
    \cmidrule{3-9}
          &       & 7  & 373     & 2.86  & 573      & 0.84  & +200      & -2.02 \\
    \cmidrule{3-9}
          &       & 8  & 20,055  & 8.03  & 12,389   & 57.66 & -7,666    & +49.63 \\
    \cmidrule{2-9}
          & Mob   & 9  & 2       & 0.00  & 0        & 0.00  & -2        & 0.00 \\
    \cmidrule{2-9}
          & Ent   & 10 & 7,161   & 18.65 & 856      & 1.16  & -6,305    & -17.49 \\
    \midrule
    \multirow{5}[5]{*}{CN} & \multirow{3}[3]{*}{Bb} & 11 & 843,375 & 39.74 & 778,618  & 24.52 & -64,757   & -15.22 \\
    \cmidrule{3-9}
          &       & 12 & 1,003,635 & 78.84 & 255,411 & 9.82  & -748,224  & -69.02 \\
    \cmidrule{3-9}
          &       & 13 & 3,877,512 & 52.99 & 2,341,134 & 17.27 & -1,536,378 & -35.72 \\
    \cmidrule{2-9}
          & \multirow{2}[2]{*}{Mob} & 14 & 190   & 0.01  & 22       & 0.00  & -168      & -0.01 \\
    \cmidrule{3-9}
          &       & 15 & 353     & 0.00  & 34       & 0.00  & -319      & 0.00 \\
    \midrule
    Total &       &       & 5,792,237 & 11.04 & 3,501,369 & 1.43  & -2,290,868 & -9.61 \\
    \bottomrule
    \end{tabular}

    \vspace{2mm}
    \parbox{\columnwidth}{\scriptsize
    India: 1=Reliance Jio, 2=BSNL, 3=Bharti Airtel, 4=Vodafone\quad
    US: 5=Comcast, 6=AT\&T, 7=Charter, 8=CenturyLink, 9=AT\&T, 10=Mediacom\quad
    China: 11=Telecom, 12=Unicom, 13=Mobile, 14=Unicom, 15=Mobile\quad
    Bb: Broadband, Mob: Mobile, Ent: Enterprise.
    }
\end{table}

%% file: main/06_discussion.tex
\section{Discussion}

\noindent
\textbf{Findings and Comparison.}
Our measurements show that IPv6 periphery deployment has expanded substantially, but its security posture has not improved uniformly.
Compared with the 2021 baseline, service exposure has declined in several major ISP blocks, and routing loop vulnerabilities have also been reduced in the historical regions.
However, millions of devices still expose sensitive services or exhibit loop-prone forwarding behavior, and the remaining risks are highly concentrated across regions, ASNs, and vendor ecosystems.
These findings suggest that IPv6 security risks are becoming structurally concentrated and operationally heterogeneous as deployment matures.

\noindent
\textbf{Newly Exposed Vulnerabilities.}
Beyond traditional service exposure, our LLM case study highlights an emerging application-layer risk.
Locally deployed LLM tools may expose APIs or WebUI interfaces when bound to public addresses without adequate authentication.
Although such exposure is rare among IPv6 network periphery devices, the confirmed Ollama case shows that AI-facing services can appear in periphery address spaces.
These interfaces may enable unauthorized model enumeration, inference abuse, or leakage of metadata, suggesting that LLM deployment endpoints should be treated as sensitive management services rather than ordinary web applications.

\noindent
\textbf{Limitations.}
Our study has several limitations.
First, our selection of regions, ISPs, and prefixes prioritizes observable IPv6 deployment and representative ISP-managed address space, which may bias results toward denser and more active networks; thus, our results should not be viewed as a complete census of the entire IPv6 Internet.
Second, the sparse IPv6 address space means that some active devices or exposed services may remain undetected.
Third, service fingerprinting and banner-to-CVE mapping indicate potential vulnerabilities rather than confirmed exploitability, due to incomplete versions, vendor modifications, or backported patches.
These limitations motivate broader longitudinal measurement and more refined validation.

\noindent
\textbf{Ethical Considerations.}
We follow strict ethical guidelines throughout our measurement activities, prioritizing the principle of ``do no harm''~\cite{durumeric2013zmap,bailey2012menlo}.
Our probing is designed to minimize disruption, with packet transmission rates carefully controlled to reduce network load.
We notify vendors or organizations of identified vulnerabilities when appropriate to support remediation.
We also anonymize and sanitize collected data, exclude personally identifiable information from analysis and released artifacts, and use the results only to promote better operational practices and strengthen the IPv6 security ecosystem.

%% file: main/07_related_work.tex
\section{Related Work}

As IPv6 deployment expands, Internet-wide measurement of IPv6 hosts, networks, and security exposures has become increasingly important.
However, the vast and sparse IPv6 address space makes exhaustive scanning infeasible, requiring adaptive discovery strategies.
Recent studies related to our work can be grouped into four themes: Internet-wide active discovery, target generation and prefix-aware refinement, security analysis of discovered IPv6 assets, and routing- or control-plane-oriented measurement.

\noindent
\textbf{Internet-wide active discovery.}
A major line of work improves the efficiency of Internet-wide IPv6 host discovery.
Recent systems such as 6SENSE~\cite{williams20246sense}, AddrProbe~\cite{cheng2025addrprobe}, HMap~\cite{hou2026hmap}, TNet~\cite{zhao2026tnet}, and 6Seeks~\cite{yang20256seeks} address the challenge of identifying active IPv6 targets under limited probing budgets.
They exploit structural guidance such as reinforcement learning, dynamic feedback, BGP-informed search, and fine-grained pattern mining to discover active addresses, networks, or periphery devices.
Related efforts also discover IPv6 router interfaces and topology-related assets at scale, such as Treestrace~\cite{yang2024efficient} and 6PSTree~\cite{wen53204336pstree}, showing the importance of scalable seedless or weak-seed measurement.

\noindent
\textbf{Target generation and prefix-aware refinement.}
Another direction studies how to generate promising IPv6 targets or refine the search space during scanning.
Recent evaluations show that target generation algorithms vary substantially in responsiveness, stability, and measurement suitability~\cite{steger2023target}.
Learning-based approaches, including 6Former~\cite{liu20236former}, 6Vision~\cite{zhang20246vision}, and Gungnir~\cite{wei2025gungnir}, infer address patterns, fully responsive prefixes, or likely active regions from limited observations.
Other works improve discovery through alternative probing primitives or address sources, such as subnet-router anycast probing~\cite{koch2025scanning}, NTP-based IPv6 scanning~\cite{klopsch2025time}, and hierarchical subnet discovery~\cite{zhou2025subrecon}.
Together, these studies show that IPv6 scanning increasingly depends on prefix awareness, response guidance, and adaptive refinement, which aligns with the design philosophy of RGPS.

\noindent
\textbf{Security analysis of discovered IPv6 assets.}
Beyond host discovery, several studies extend IPv6 measurement toward security analysis.
6SENSE conducts Internet-wide scanning-driven security analysis of IPv6 hosts, covering open ports, security-sensitive services, and IPv6-specific TLS exposure~\cite{williams20246sense}.
HMap and TNet also analyze exposed internal devices, gateway-related risks, and security-sensitive services in discovered IPv6 networks~\cite{hou2026hmap,zhao2026tnet}.
Other work highlights operational risks in deployed IPv6 environments, including reflection amplification exposure~\cite{hu2025grey}, NAT-centric consumer gateway security~\cite{olson2023doomed}, insecurity among NTP-sourced consumer and IoT hosts~\cite{klopsch2025time}, and privacy leakage from IPv6 addressing practices~\cite{saidi2022one}.
These efforts show that IPv6 measurement is increasingly used for Internet-scale security assessment.

\noindent
\textbf{Routing and control-plane measurement.}
A final line of work studies routing behavior, ICMPv6 signals, and control-plane anomalies.
Maier \emph{et al.} study persistent routing loops in today’s IPv4/IPv6 Internet and show that IPv6 exhibits high threat potential~\cite{maier2023loop}.
Pan \emph{et al.} exploit ICMP rate-limiting side channels to measure IPv6 networks remotely and infer source address validation and reachability~\cite{pan2022your}.
Holzbauer \emph{et al.} analyze what ICMPv6 error messages reveal about their sources, highlighting the measurement value of protocol-level feedback~\cite{holzbauer2024destination}.
Longitudinal observations of IPv6 scanners further show that scanning behavior is shaped by BGP announcements and prefix visibility, emphasizing prefix-level measurement bias~\cite{egloff2025detailed}.

Overall, prior work has made substantial progress in IPv6 target generation, active discovery, topology inference, and security measurement.
However, existing studies usually focus on a single layer, such as active address discovery, router-interface collection, or specific exposure classes, and have paid limited attention to IPv6 network periphery devices as a distinct security target.
They also lack a lightweight strategy for quickly selecting high-value BGP-announced prefixes before large-scale probing.
In contrast, our work revisits IPv6 network peripheries from an integrated perspective, combining prefix-level selection, large-scale periphery discovery, service exposure analysis, LLM-related unauthorized-access case study, and routing-loop validation to provide an updated global security view of the contemporary IPv6 edge.

%% file: main/08_conclusion.tex
\section{Conclusion}

This paper revisits and expands the measurement of IPv6 network peripheries through RGPS-based global scanning, identifying 281.9M active devices with highly uneven distribution across regions and networks.
Based on this dataset, we analyze exposed services, LLM deployment-tool exposure as a service-exposure case study, and routing loop vulnerabilities, showing that security risks persist despite the rapid growth of IPv6 deployment.
Overall, our findings provide an updated empirical view of the contemporary IPv6 periphery and highlight the need for continuous measurement and mitigation of exposed services and routing anomalies.

% \section{Conclusion}

% In this paper, we revisit and expand the measurement of IPv6 network peripheries at a global scale.
% Using the proposed Response-Guided Prefix Selection (RGPS) strategy, we identify a large population of active IPv6 network periphery devices and characterize their highly uneven global distribution.

% Based on this measurement, we further examine three major security issues in the contemporary IPv6 periphery: exposed services, unauthorized access risks in LLM deployment tools, and routing loop vulnerabilities.
% Our results show that, despite the continued growth of IPv6 deployment, important security risks remain present in multiple forms and across multiple layers.

% Overall, this study suggests that the security challenges of the IPv6 periphery have not disappeared with deployment growth, but instead continue to evolve alongside operational practices, software ecosystems, and emerging services.
% We hope that our measurement framework and findings can support future longitudinal studies and contribute to the continuous improvement of the IPv6 security ecosystem.

%% file: main/09_acknowledgement.tex
\section*{Acknowledgment}
Authors were supported by the National Natural Science Foundation of China (No. 62502236, No. U25B2025), the Natural Science Foundation of Tianjin (No. 24JCQNJC02070), and the Open Project of Key Laboratory of Industrial Information Security Perception and Evaluation Technology, Ministry of Industry and Information Technology.